\documentclass[11pt]{article}
\usepackage{lscape}
\usepackage{subcaption}
\usepackage{authblk} 
\usepackage[a4paper, total={7in, 10in}]{geometry}
\usepackage{xcolor} 
\usepackage{xr-hyper}
\usepackage{hyperref}

\usepackage[super,sort&compress]{natbib} 
\usepackage{amssymb,amsmath,amsthm,amsfonts,physics,mathptmx,cuted,esdiff}

\usepackage{times}
\usepackage{bm}
\usepackage[dvips]{epsfig}
\usepackage{graphicx}
\usepackage{psfrag}
\usepackage{amssymb}
\usepackage{caption}
\usepackage{subcaption}
\usepackage{amsmath,xcolor}

\usepackage{color}

\newcommand{\rd}{\textcolor{black}}

\newcounter{remcount}

\newcounter{excount}

\newtheorem{remm}[remcount]{Remark}

\newtheorem{ex}[excount]{Example}

\begin{document}
\title{Graded lithium-ion battery pouch cells to homogenise current distributions and \rd{reduce} lithium plating}


\date{}
\author[1]{R.~Drummond\thanks{Email: \texttt{ross.drummond@sheffield.ac.uk}}}
\author[2,3]{E.~C.~Tredenick}
\author[4]{T.~L.~Kirk}
\author[3,5]{M.~Forghani}
\author[3,5]{P.~S.~Grant}
\author[2,3]{S.~R.~Duncan}

\affil[1]{\small School of Electrical and Electronic Engineering, University of Sheffield, Sheffield, S1 3JD, UK}
\affil[2]{Department of Engineering Science, University of Oxford, Oxford, OX1 3PJ, UK}
\affil[3]{The Faraday Institution, Quad One,  Harwell Campus, Didcot, OX11 0RA, UK}
\affil[4]{Department of Mathematics, Imperial College London, London, SW7 2AZ, UK}
\affil[5]{Department of Materials, University of Oxford, Oxford, OX1 3PH, UK}

\maketitle

\begin{abstract}
Spatial distributions in current, temperature, state-of-charge and degradation across the plane of large format lithium-ion battery pouch cells can significantly impact their performance, especially at high C-rates. In this paper, a method to smooth out these spatial distributions by grading the electrode microstructure in-the-plane is proposed. A mathematical model of a large format pouch cell is developed and validated against both temperature and voltage experimental data. An analytical solution for the optimal graded electrode that achieves a uniform current distribution across the pouch cell is then derived. The model predicts that the graded electrodes could significantly reduce the likelihood of lithium plating in large format pouch cells, with grading increasing the C-rate at which plating occurs from 2.4C to 4.3C. These results indicate the potential of designing spatially varying electrode architectures to homogenise the response of large format pouch cells and improve their high rate performance.
\end{abstract}
\section{Introduction}
Lithium-ion battery electrodes are complex non-linear dynamical systems reacting non-uniformly in space and time, especially when subject to high C-rate currents.  Phenomena such as travelling wave fronts in LiFePO$_4$ (LFP) cathodes during charging \cite{singh2008intercalation,tredenick2024multilayer} and localised lithium-plating in graphite anodes \cite{ho20213d} are just some examples of the diverse range of non-linear behaviours that have been observed in battery electrodes, and it is these localised behaviours that often govern cell performance. To protect against these spatially localised electrochemical behaviours, the use of graded electrode structures (as in those with controlled spatial variations in electrode microstructure) has been proposed with the idea being that introducing heterogeneity into the electrode microstructure can reduce the heterogeneity in the electrochemical response.

Recently, experimental data have provided support for the use of graded electrodes in practical cells \cite{lauro2023restructuring}. A common theme is the potential of graded electrodes to increase both the capacity at high C-rates \cite{sukenik2023impact, chowdhury2021simulation}  and the cycle life  of thick electrodes \cite{cheng2022extending,cheng2020combining}, with other benefits including a   reduction in electrode tortuosity \cite{chen2024tortuosity}. Data for both standard electrodes, such as LFP cathodes \cite{cheng2022extending,cheng2020combining}, and non-standard electrodes, such as anodes arranged in a layered structure composed of high power Li$_4$Ti$_5$O$_{12}$ and high capacity SnO$_2$ \cite{lee2021multi}, have shown similar benefits. However, care has to be taken when designing graded electrodes, as structuring the electrode incorrectly can deteriorate performance \cite{yari2020constructive, cheng2020combining}. To prevent this, mathematical models have been developed to predict the electrochemical response of graded electrodes so as to quickly and cost-effectively optimise their designs. Several different models have now been developed covering a range of length scales (including micro-structural\cite{boyce2021design,tredenick2023bridging} and  Doyle-Fuller-Newman-type models\cite{wu2022gradient,tredenick2024multilayer}) and performance metrics (such as fast charging capacitance \cite{tredenick2024multilayer} and electrical impedance \cite{drummond2022modelling}). These results highlight the growing maturity of battery modelling and parameterisation, which in turn, is making it increasingly straightforward to embed modelling pro-actively in the battery design process \cite{garrick2023atoms}. An example of this battery co-design process are the trapezoidal electrodes\cite{cheng2022extending} that were developed using insights \cite{drummond2022modelling} on how the distribution of carbon and binder affects the conductivity of the LiFePO$_4$ electrodes.


The focus of most existing studies on graded electrodes for lithium-ion batteries has been on controlling the \textit{through-thickness} local electrode microstructure \cite{boyce2021design, ramadesigan2010optimal}, including  bilayer cathodes\cite{tredenick2024multilayer} composed of discrete layers of both $\quad$ Li[Ni$_{0.6}$Co$_{0.2}$Mn$_{0.2}$]O$_{2}$ (NMC622) and LFP. Focusing on the through-thickness electrode structure (with experiments typically only involving coin cells \cite{cheng2022extending}) is justified by the fact that it is charge mobility through the electrode thickness between the anode and cathode that governs coin cell performance. However, in recent years, there has been a general trend in commercial cells towards larger cell formats (demonstrated, for example, by the introduction of the larger 4680 cylindrical cells \cite{li2023enable}). The idea being that larger cell formats increase the specific energy density, as the ratio of active material to superfluous materials such as casings and control circuits is increased. Large format cells have their own problems compared to coin cells; in particular,  they experience new forms of spatial heterogeneities, which emerge at the macro-scale (as in across their width and length).  In particular, heterogenous state-of-charge  \cite{liu2010visualization}, thermal hot spots \cite{chu2020parameterization,lin2022multiscale} and spatially localised distributions of electrode degradation \cite{mikheenkova2024visualizing, fordham2023correlative,yari2022non}, lithium plating \cite{yang2019asymmetric}, and currents in-the-plane of large-format pouch cells \cite{bason2022non,mohammadi2019diagnosing} have been identified as impacting performance, especially in high rate applications. 

The significance of in-the-plane spatial heterogeneities on the performance of large format pouch cells is becoming increasingly apparent, and has driven efforts to understand how these heterogeneities develop using mathematical models. Through these models, predictions for chemical reaction \cite{wang2024plane} and current distributions along the plane of the current collectors  \cite{yazdanpour2014distributed} (and how these distributions evolve with changes in the cell's design,  e.g. its current collector thickness \cite{campillo2017effect} and tapering \cite{cho2023improving}) have been developed, with analytical solutions even being proposed \cite{taheri2014theoretical}. The push to understand the planar dynamics of pouch cells has meant that the number of spatial dimensions in these models has increased, with 2D \cite{newman1993potential} and 3D \cite{parmananda2023underpinnings, hahn2023reduced,guo2013three,pan2020computational,lin2022multiscale,aylagas2022cidemod} effects often included, as opposed to the through-thickness models of, for example, the DFN model and single particle models \cite{doyle1993modeling,santhanagopalan2006review}.  One of the main drawbacks of large format battery models is their complexity, which is a consequence of their multi-dimensional spatial domains. This model complexity leads to several challenges including long simulation times and difficulties in model parameterisation, as discussed in the electro-thermal modelling studies \cite{chu2020parameterization}\cite{lin2022multiscale}. These challenges demonstrate a need for easily parameterised and computationally efficient pouch cell models able to capture the spatial distributions seen in experimental data. This problem is considered with the model introduced in Section \ref{sec:Model}.  

\begin{figure}
     \centering
     \begin{subfigure}[t]{0.3\textwidth}
         \centering
         \includegraphics[width=0.8\textwidth]{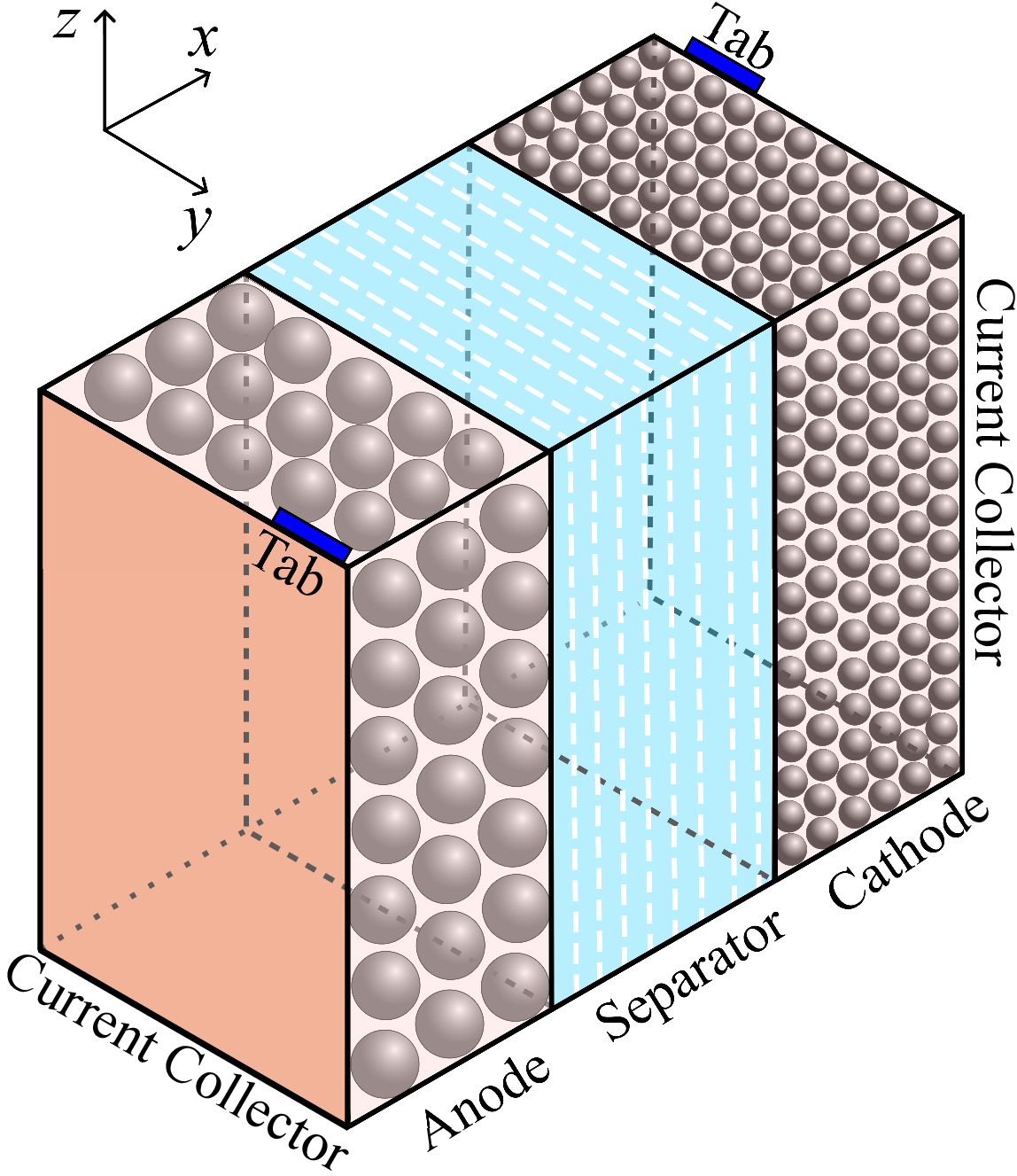}
         \caption{\rd{Representation of electrode/separator sandwich of the pouch cell. A left handed co-ordinate system is used. }}
         \label{fig:sandwich}
     \end{subfigure}
     \hfill
     \begin{subfigure}[t]{0.3\textwidth}
         \centering
         \includegraphics[width=0.8\textwidth]{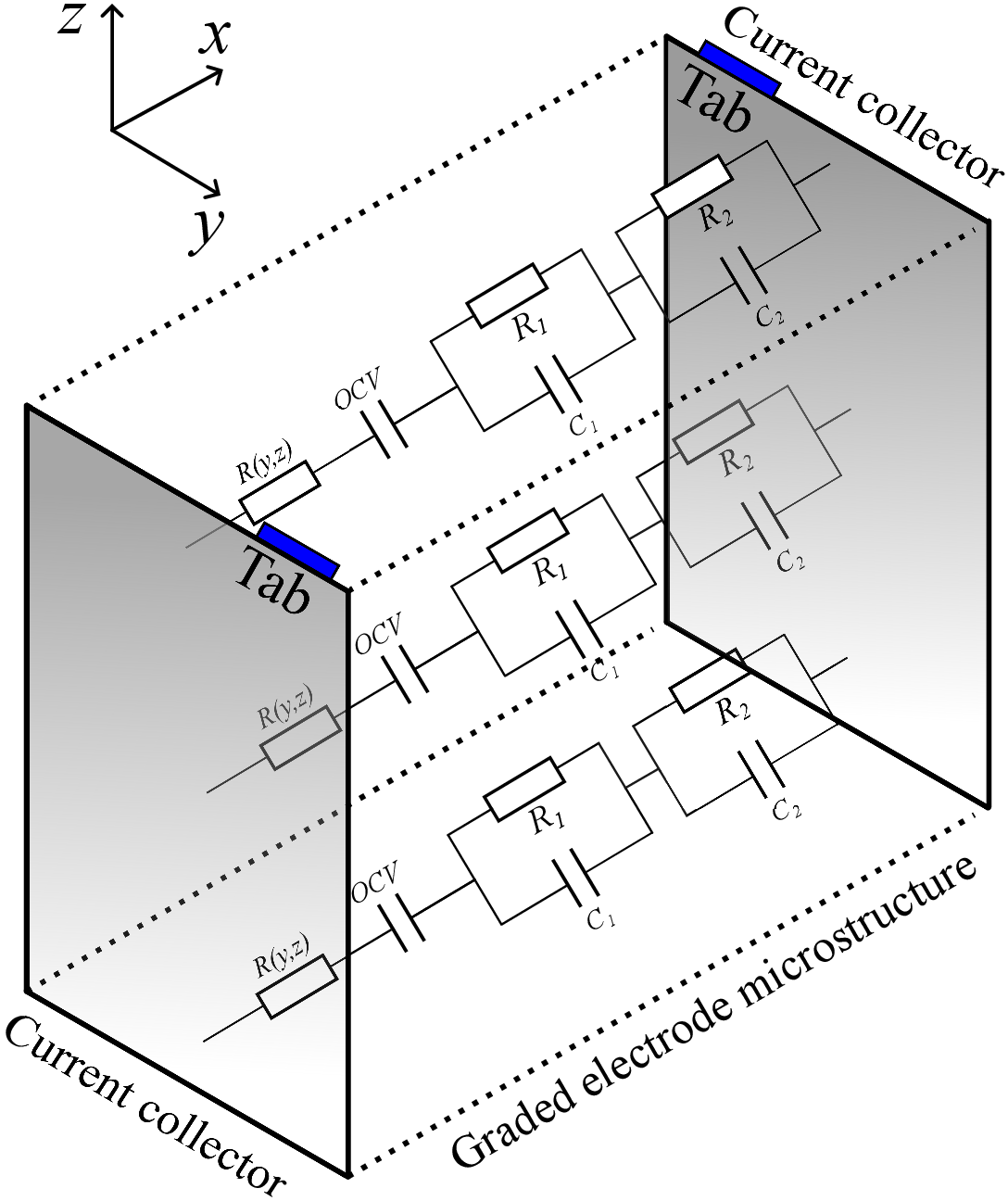}
         \caption{Model simplification for large format pouch cells with the electrode thickness being significantly thinner than the width and length. Circuit parameters are found in Table \ref{tab:params}.}
         \label{fig:assumption}
     \end{subfigure}
     \hfill
     \begin{subfigure}[t]{0.3\textwidth}
         \centering
         \includegraphics[width=\textwidth]{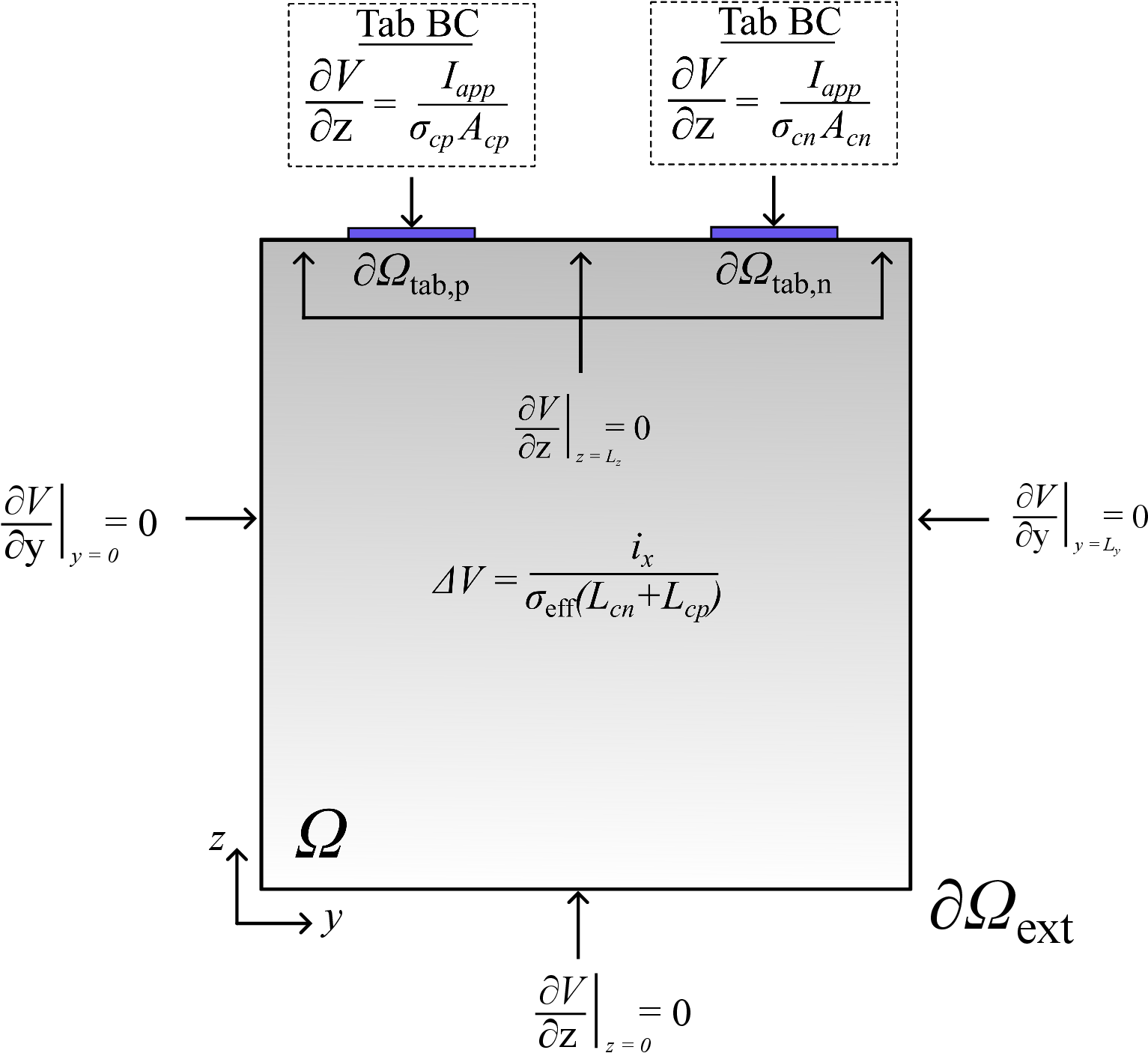}
         \caption{Spatial domain of the 2D pouch cell model.  }
         \label{fig:domain}
     \end{subfigure}
        \caption{Schematic of the 2D model for a lithium-ion battery pouch cell, \rd{Figure \ref{fig:sandwich} is approximated with Figure \ref{fig:assumption} and Figure \ref{fig:domain}. The model is a combination of Figure \ref{fig:assumption} and Figure \ref{fig:domain}.}  }
        \label{fig:pouch_model}
\end{figure}

The formation of spatial heterogeneities across pouch cells indicates that it may be possible to grade the local electrode microstructure \textit{in-the-plane} to mitigate their effect, similar to what has been achieved with through-thickness grading \cite{hosseinzadeh2018impact,ramadesigan2010optimal}. The graded electrode problem is illustrated in Figure \ref{fig:pouch_model}, with Figure \ref{fig:sandwich} indicating the planar and through-thickness directions of the pouch cell, \rd{Figure \ref{fig:assumption} shows the simplification of the model to a 2D problem with RC circuits distributed across the plane}, and Figure \ref{fig:domain} illustrates the 2D model domain. \rd{Figure \ref{fig:sandwich} is approximated with Figure \ref{fig:assumption} and Figure \ref{fig:domain}. The model is a combination of Figure \ref{fig:assumption} and Figure \ref{fig:domain}.}

\paragraph{Contributions:} In this paper, the problem of modelling and optimising graded electrode microstructures for large format pouch cells with controlled variations in resistance in-the-plane is considered. 
 The following contributions are obtained:
\begin{enumerate}
    \item A model for a large format pouch cells is described and validated against experimental data\cite{lin2022multiscale}. 
    \item An analytical solution for the distribution of carbon black in the electrode plane that achieves a flat current distribution across the pouch cell is derived.
    \item Compared to uniform electrodes, it is predicted that graded electrodes can increase the charging rate at which lithium plating occurs from 2.4C to 4.3C.
\end{enumerate}
It is proposed that a primary advantage of these graded pouch cells is to mitigate the impact of lithium plating (see Section \ref{sec:plating}), as the flatter current distributions reduce the likelihood of negative anode overpotentials. The results highlight the importance of in-the-plane effects when characterising lithium plating in large format cells. By contrast, it is shown that the high thermal/electronic conductivity of the current collectors means that only minimal benefits of grading are observed when considering the thermal and voltage response; Section \ref{sec: improvement} shows that the graded cells only delivered a 1.2\% increase in stored capacitance  for 4C charging compared to the uniform electrodes, a conclusion in agreement with Hosseinzadeh et al.,~2018\cite{hosseinzadeh2018impact}.  These validated modelling results indicate the potential of grading to improve the cycle-life and high-rate performance of large format pouch cells. More broadly, they indicate how improvements in battery design and manufacturing could reduce the performance gap observed between coin cells constructed in laboratories and the large format cells used in commercial high-rate applications \cite{frith2023non}.

	\section{Mathematical model }\label{sec:Model}

	In this section, a mathematical model of a large format pouch cell is introduced. Specific focus is given to a comparison between the model and experimental data\cite{lin2022multiscale}, with the presented model being simpler in form (specifically, here, circuit dynamics are used to capture the through-thickness electrical response, whereas a simple electrochemical model is used in Lin et al.,~2022\cite{lin2022multiscale}) whilst still being able to accurately capture the in-plane spatial distributions seen in the experimental data (see Section \ref{sec:exp_comp}). Keeping the model formulation relatively simple while retaining the important physics-based mechanisms, allows analytical solutions to the optimal graded electrode design problem to be obtained (explored in Section \ref{sec:grading}), which is the focus of this work.


  \subsection{Model formulation}
Figure \ref{fig:pouch_model} presents a schematic for the model's formulation and spatial domain.  The model parameters and variables are defined in Tables \ref{tab:vars} and \ref{tab:params}.  The following assumptions are made:
\begin{enumerate}
\item[\textbf{A1}:] The thicknesses of the current collectors are much thinner than their width and height.
\item[\textbf{A2:}] The circuit models of Figure \ref{fig:assumption} describe the electrical dynamics at each point in the plane of the pouch cell.
\item[\textbf{A3:}] The multi-layer pouch cell can be modelled as an effective single layer\cite{chu2020parameterization,lin2022multiscale}.
\end{enumerate}
\textbf{A1} is justified because large format pouch cells are considered (with typical current collector thicknesses being $\approx 10 \mu$m and their planar dimensions being $\approx$ 15cm $\times$ 20cm).  Defining $i_x(y,z,t)$ as the current density through the electrode, and $I_{x,k}(x,y,z,t)$ as the current density flowing in the $x$-direction through current collector $k$, then, under this thin plate assumption, the current density gradient in the $x$-direction through the current collectors can be approximated by
\begin{align}
\frac{\partial I_{x,k}(x,y,z,t)}{\partial x} \approx \pm \frac{i_x(y,z,t)}{L_{ck}}, \quad k \in \{n,p\},
\end{align}
following the ideas of Taheri et al.,~2014\cite{taheri2014theoretical}.
\textbf{A2} is a simplification of the electrode electrical dynamics, but, even so, the resulting model is able to capture the spatial distributions seen in the experimental data (see Section \ref{sec:exp_comp}) while being simple enough to generate an analytical solution to the optimal electrode design problem (\rd{see }Section \ref{sec:optimisation}). \textbf{A3} is justified under the assumption that each layer has a similar composition and so reacts similarly. 


%
%

\subsection{Model equations}
The following equations characterise the proposed mathematical model for large-format pouch cells, with the models domain described in Figure \ref{fig:domain}. Regarding the notation, each model variable in Eqn. \eqref{model_eqns} is defined in the ($y,z$)-plane of the pouch cell, as illustrated in Figure \ref{fig:domain}, and may also evolve in time $t$. This $(y,z,t)$ dependency of the variables is dropped in the model equations \eqref{model_eqns} to ease readability. $V(y,z,t)$ is the voltage (defined as the potential of the current collector at the cathode minus that of the anode) at each point in the plane, $T(y,z,t)$ is the cell temperature, $i_x(y,z,t)$ is the local through-thickness current density, $R(y,z)$ is the local resistance of the two electrodes and the separator that is constant with time, while SoC$(y,z,t)$ is the local state of charge in the plane.  Table \ref{tab:vars} summarises the model's variables, while Table \ref{tab:params} gives its parameters. For the graded cell, the only parameter varying in $(y,z)$ is $R(y,z)$, with the rest being constants. Using the notation $\Delta f = \frac{\partial^2 f}{\partial y^2} + \frac{\partial^2 f}{\partial z^2}$ and defining
	\begin{align}
\sigma_\text{eff}(L_\text{cn}+L_\text{cp})& =     \dfrac{\  L_{\scriptscriptstyle \text{cn}} L_{\scriptscriptstyle \text{cp}} \ \sigma_{\scriptscriptstyle \text{cn}} \  \sigma_{\scriptscriptstyle \text{cp}}  }{ \sigma_{\scriptscriptstyle \text{cn}}  L_{\scriptscriptstyle \text{cn}} + \sigma_{\scriptscriptstyle \text{cp}}   L_{\scriptscriptstyle \text{cp}}  }  ,  \label{eq11}  
	\end{align}
 then
 \begin{subequations}\label{model_eqns}
	\begin{align}
    i_x  & =  \frac{ V  - U(\text{SoC,T})-v_1 -v_2   }{   
    R\, A_{\scriptscriptstyle \text{cell}}},   \label{eq1} \\
\Delta V   &=  \frac{  i_x}{ \sigma_\text{eff}(L_\text{cn}+L_\text{cp})   }    ,\ \   \text{on} \ \   \Omega, \ t>0, \label{eq2}  \\
\rho \frac {\partial T}{\partial t} &=  \lambda_{\scriptscriptstyle \text{eff}}\;\Delta T  + \dfrac{1}{L_x \ n_\text{layers}}  \left[R\;A_{\scriptscriptstyle \text{cell}} \;{i_x}^2 -h_{\text{faces}}\left(T-T_{\text{ref}} \right) \right]         
,\ \  \text{on} \ \   \Omega    , \ t>0, \label{eq3}  \\
 \frac{\partial \text{SoC}} {\partial t} & = \rd{\frac{    A_{\scriptscriptstyle \text{cell}}}{  C_{\scriptscriptstyle \text{cell}}   }i_x},\ \  \text{on} \ \  \Omega, \ t>0, \label{eq4}  \\
\frac{\partial v_1}{\partial t} & =- \frac{v_1 }{R_1C_1} +\rd{ \frac{A_{\scriptscriptstyle \text{cell}}}{C_1}i_x}   , \label{eq5}\\
\frac{\partial v_2}{\partial t} & =- \frac{v_2}{R_2C_2} +\rd{\frac{A_{\scriptscriptstyle \text{cell}}}{C_2} i_x }  ,\label{eq6}
	\end{align}
\end{subequations}
where $\Omega = [0,\;L_y] \times [0,\;L_z]$ is the projection of the cell onto the $(y,z)$-plane and $L_y$ and $L_z$ are the cell width and height, respectively. \rd{The physical meaning of Eqns. \eqref{eq5} - \eqref{eq6} is as follows. Eqn. \eqref{eq1} is obtained from  the circuit models\cite{plett2015battery} distributed across the plane of the pouch cell in Figure \ref{fig:assumption}. Eqn. \eqref{eq1} defines the through-thickness current density (as in the current density flowing across the electrodes charging the active material particles). Eqn. \eqref{eq2} is adapted from Taheri et al.,~\cite{taheri2014theoretical} to account for the circuit dynamics and defines the distribution of the voltage in the $(y,z)$ plane, given the current density $i_x$ through the cell at each point. Eqn. \eqref{eq3} defines the localised temperature dynamics of the cell and is adapted from Lin et al.,~\cite{lin2022multiscale} with Joule heating defined from Eqn. \eqref{eq1} instead of electrochemical potentials\cite{lin2022multiscale}. Using the circuit model of Figure \ref{fig:assumption} to describe the electrical dynamics at each point ($y,z)$ in the plane (with the derivations of these dynamics discussed in Plett\cite{plett2015battery}), Eqn. \eqref{eq4} describes the state-of-charge and Eqns. \eqref{eq5} - \eqref{eq6} are the relaxation dynamics associated with Figure \ref{fig:assumption}'s RC pairs.}

\subsection{Boundary Conditions}
Using the notation of $n_{\perp}$ as the two-dimensional unitary vector normal to the boundary and $\nabla T(y,z,t) = [\partial T/\partial y,\,\partial T/\partial z]^\top$, then the model's boundary conditions are:
	\begin{subequations}\label{eqn:bcs_all}
		\begin{align}
	n_{\perp} \cdot	\nabla T
  & = -\frac{h}{\lambda_{\scriptscriptstyle \text{eff}}} \left( T - T_{\scriptscriptstyle \text{ref}} \right)  ,\ \  \text{on} \ \  \partial \Omega_{\scriptscriptstyle \text{ext}}  , \ t>0, \label{bveq1}
  \\
		n_{\perp} \cdot	\nabla T & = -\frac{h_{\scriptscriptstyle \text{tab}}}{ \lambda_{\scriptscriptstyle \text{eff}}}  \left( T - T_{\scriptscriptstyle \text{ref}} \right)  ,\ \  \text{on} \ \  \partial \Omega_{\scriptscriptstyle \text{tab}} , \ t>0, \label{bveqtabs}
  \\
		\diff {\text{V}}z     	 \Big|_{z = L_z}    	& = \frac{ 
 I_{\scriptscriptstyle \text{app}}   }{ \sigma_{\scriptscriptstyle \text{ck}}   	A_{\scriptscriptstyle \text{tab}}}   ,\ \  \text{on} \ \  \partial \Omega_{\scriptscriptstyle \text{tab,k}} , ~k \in \{n,\,p\},\, t>0, \label{bveq2}  
\\
	n_{\perp} \cdot \nabla V	& =    0   ,\ \  \text{on} \ \  \partial \Omega_{\scriptscriptstyle \text{ext}} , \ t>0, \label{bveq4}
		\end{align}
	\end{subequations}
where $\partial \Omega_{\scriptscriptstyle \text{tab,n  }}$ and $\partial \Omega_{\scriptscriptstyle \text{tab,p}}$ are the negative and positive tabs, $\partial \Omega_{\scriptscriptstyle \text{tab  }}= \partial \Omega_{\scriptscriptstyle \text{tab,n  }}\cup \partial \Omega_{\scriptscriptstyle \text{tab,p  }}$, $\partial \Omega$ is the external boundary region and $\partial \Omega_{\scriptscriptstyle \text{ext}}$ is the external boundary region without the tabs, $\partial \Omega_{\scriptscriptstyle \text{ext}} = \partial \Omega \setminus \partial \Omega_{\scriptscriptstyle \text{tab}}$.

\subsection{Numerical Solution Procedure}\label{sec:NumSol} 
To simulate the model, the partial-differential-algebraic equations of Eqns. \eqref{model_eqns} and \eqref{eqn:bcs_all} were discretised in space using Chebyshev spectral collocation\cite{trefethen2000spectral} and solved using “\texttt{ode15s}” in MATLAB\textregistered~2022b\cite{Matlab2022}. On average, a 5C charge of a uniform cell took 12.7~s (with 24x24 nodes) when run on a desktop Macintosh (MacBook Pro, Monterey, v~12.4, M1 2020, 16GB, 256GB, 64 bit). The unknown model parameters were found by fitting to 4C square-wave-excitation experiments starting at 30\% SOC \cite{lin2022multiscale}. The open-source code for the pouch cell model simulations can be found at: \url{https://github.com/EloiseTredenick/2DBatteryTemperatureGradedModel}.


\paragraph{Comparison to existing models:} The main differences between the large-format pouch cell model of \eqref{model_eqns} and existing models \cite{lin2022multiscale} \cite{taheri2014theoretical} are: \textit{i}) the inclusion of the spatially-varying resistance, $R(y,z)$, in \eqref{model_eqns},  \textit{ii}) the model's relative simplicity, with no DFN-type through thickness electrochemical modelling being used; instead, the circuit of Figure \ref{fig:assumption} is distributed in the plane. The relative simplicity of the model is a deliberate design choice as it significantly simplifies the model parameterisation problem, which is crucial for capturing the experimental data. 

 It is also noted that graded planar cells were also investigated in Hosseinzadeh et al.,~2018\cite{hosseinzadeh2018impact}, where a DFN-type model with grading both in-the-plane and through-the-thickness of the electrodes was developed. The two main differences between this paper and Hosseinzadeh et al.,~2018\cite{hosseinzadeh2018impact} are: \textit{i}) an analytical solution for planar graded electrodes to achieve a uniform current distribution is obtained here (see Eqn. \eqref{eqn:sol}),  \textit{ii}) in this paper the model predictions are validated against experimental data from Lin et al.,~2022\cite{lin2022multiscale}, \textit{iii}) it is proposed that the benefits of in-the-plane grading are to reduce electrode degradation (especially plating), whereas voltage and thermal considerations were considered in \cite{hosseinzadeh2018impact}.  


\subsection{Comparison to experimental data} \label{sec:exp_comp}

The experimental data from Lin et al.,~2022\cite{lin2022multiscale} is used to validate the lithium-ion battery pouch cell model of \eqref{model_eqns}. Briefly, the data is obtained from square-wave-excitation cycling experiments starting at 30\% state of charge (SoC), using 20 Ah pouch cells from A123 Systems with LFP positive electrodes and graphite negative electrodes, which were cycled at 4C (at currents $\pm 80$ A) \rd{for} 100 seconds. A thermal imaging camera was then used to capture thermograms of the cell surface and to visualise the cell's temperature distribution across the plane during the cycling.

\subsection{Model validation}
\begin{figure}
     \centering
     \begin{subfigure}[t]{0.46\textwidth}
         \centering
         \includegraphics[width=\textwidth]{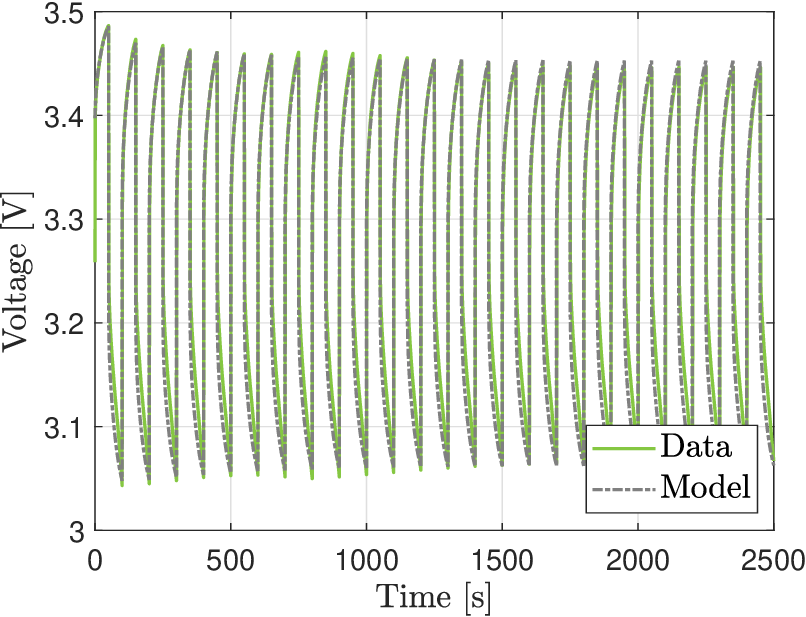}
         \caption{Comparison between the average voltages of the experimental data and the model of \eqref{model_eqns} during cycling at 4C.}
         \label{fig:valid_volt}
     \end{subfigure}
     \hfill
 \begin{subfigure}[t]{0.46\textwidth}
         \centering
         \includegraphics[width=\textwidth]{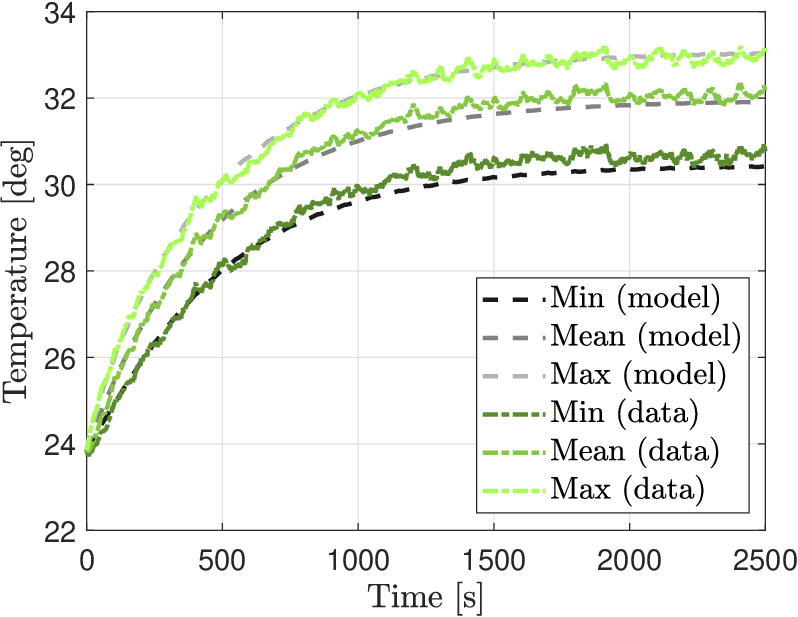}
                \caption{Comparison of the minimum, average, and maximum temperatures across the plane of the pouch cell between the experimental data and the model of \eqref{model_eqns} at 4C.}
         \label{fig:valid_temp}
     \end{subfigure}
        \caption{Comparison between the pouch cell temperature and voltage responses of the mathematical model of \eqref{model_eqns} and the experimental data of Lin et al.,~2022\cite{lin2022multiscale}.}
        \label{fig:valid}
\end{figure}

\begin{figure}
     \centering
     \begin{subfigure}[t]{0.41\textwidth}
         \centering
         \includegraphics[width=\textwidth]{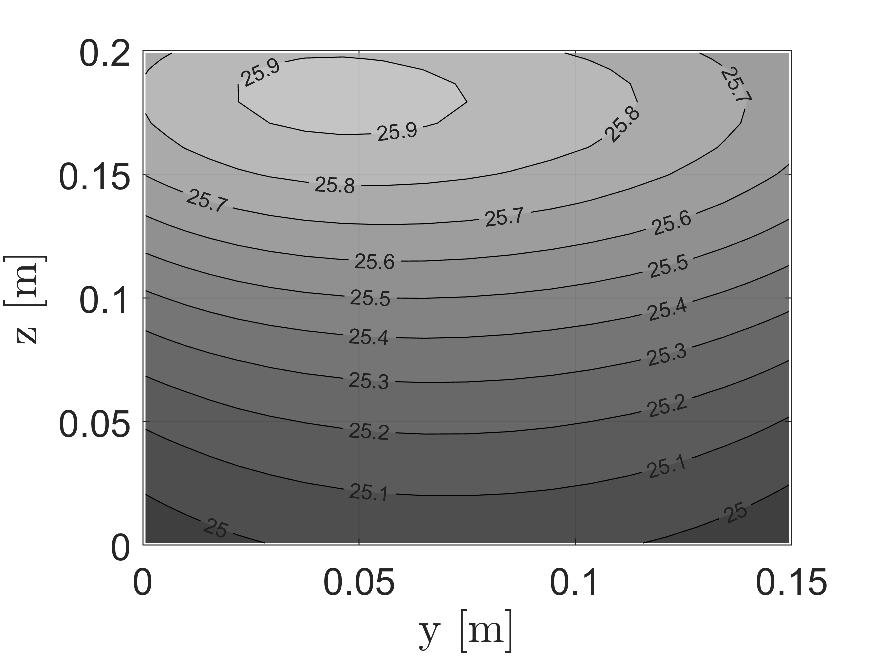}
                \caption{Model: $t = 100 $s. }   
     \end{subfigure}
\hspace{1cm}
 \begin{subfigure}[t]{0.41\textwidth}
         \centering
         \includegraphics[width=\textwidth]{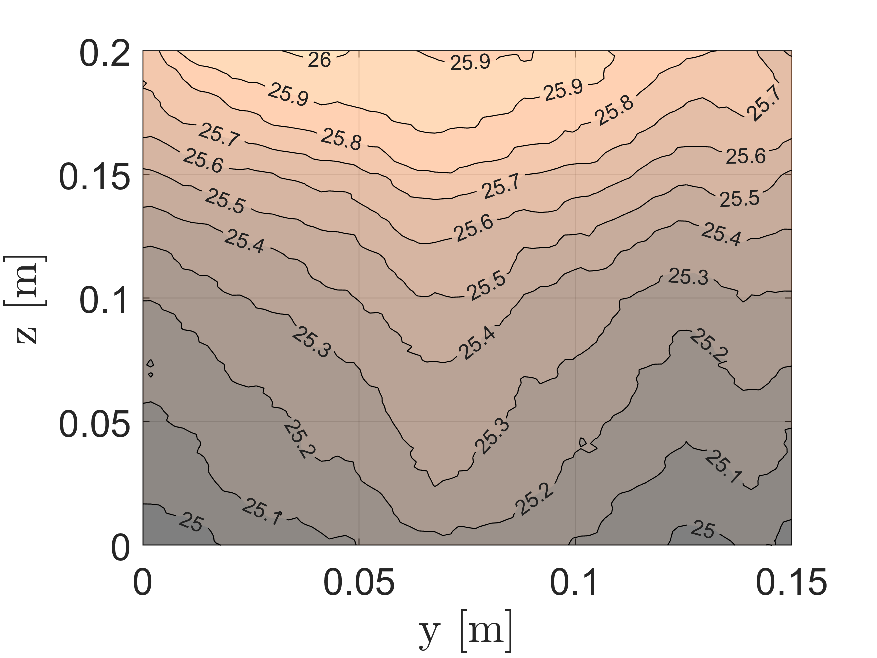}
                \caption{Data: $t = 100 $s. }
     \end{subfigure}
     \begin{subfigure}[t]{0.41\textwidth}
         \centering
         \includegraphics[width=\textwidth]{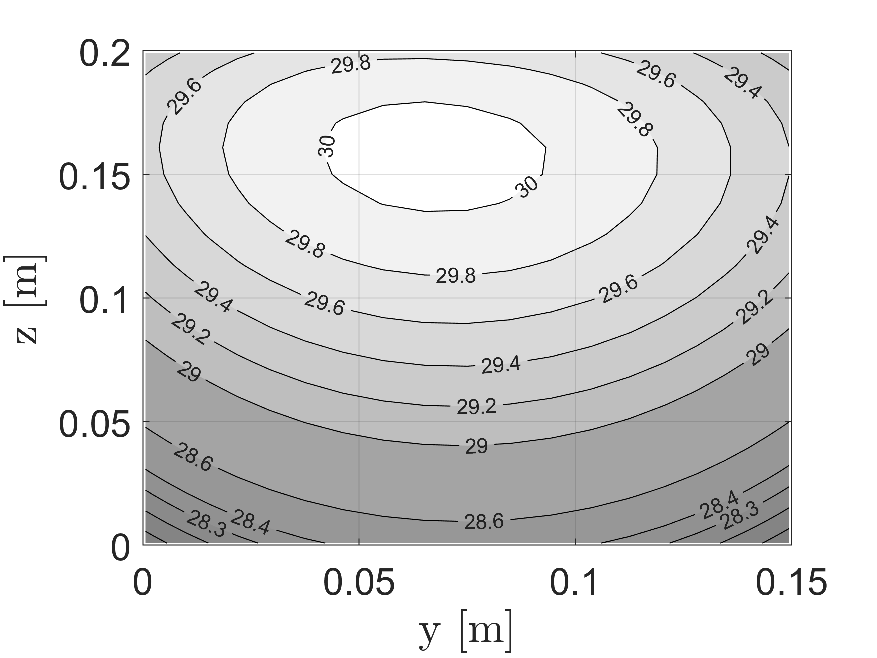}
                \caption{Model: $t = 500 $s. }
     \end{subfigure}
     \hspace{1cm} 
     \begin{subfigure}[t]{0.41\textwidth}
         \centering
         \includegraphics[width=\textwidth]{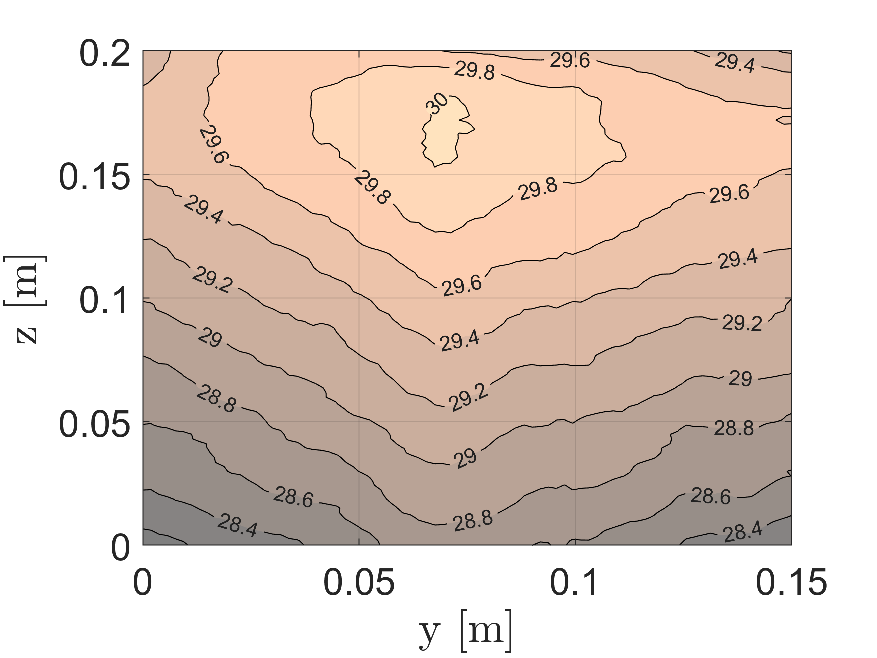}
       \caption{Data: $t = 500 $s. }
     \end{subfigure}
     \begin{subfigure}[t]{0.41\textwidth}
         \centering
         \includegraphics[width=\textwidth]{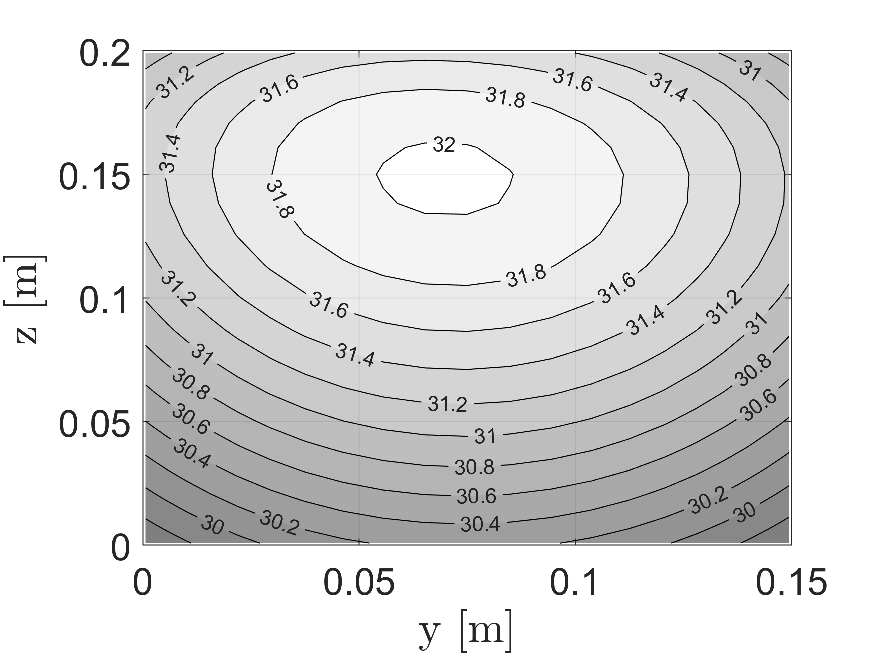}
                \caption{Model: $t = 1000 $s. }
     \end{subfigure}
    \hspace{1cm}
     \begin{subfigure}[t]{0.4\textwidth}
         \centering
         \includegraphics[width=\textwidth]{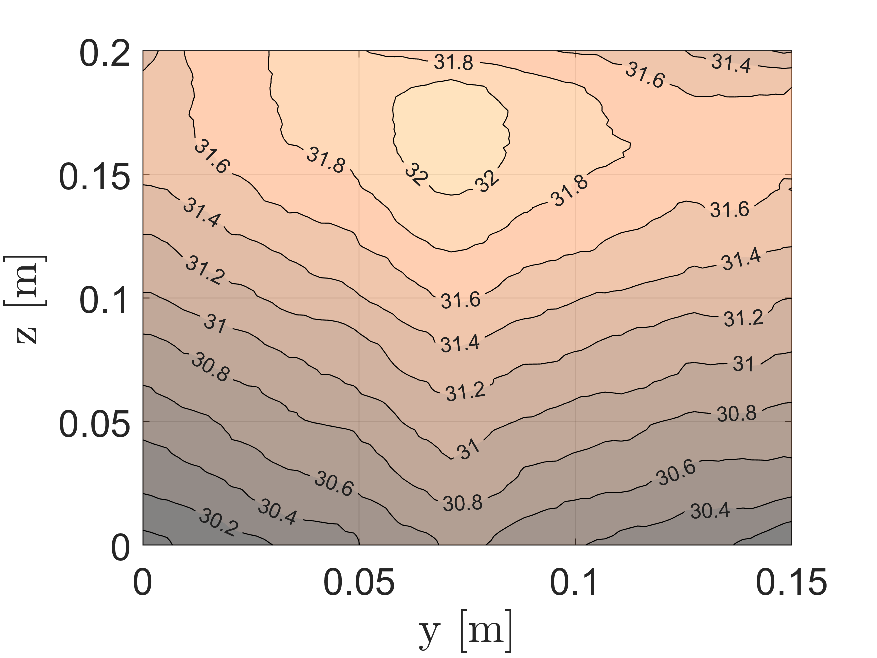}
         \caption{Data: $t = 1000 $s. }
     \end{subfigure}
     \begin{subfigure}[t]{0.4\textwidth}
         \centering
         \includegraphics[width=\textwidth]{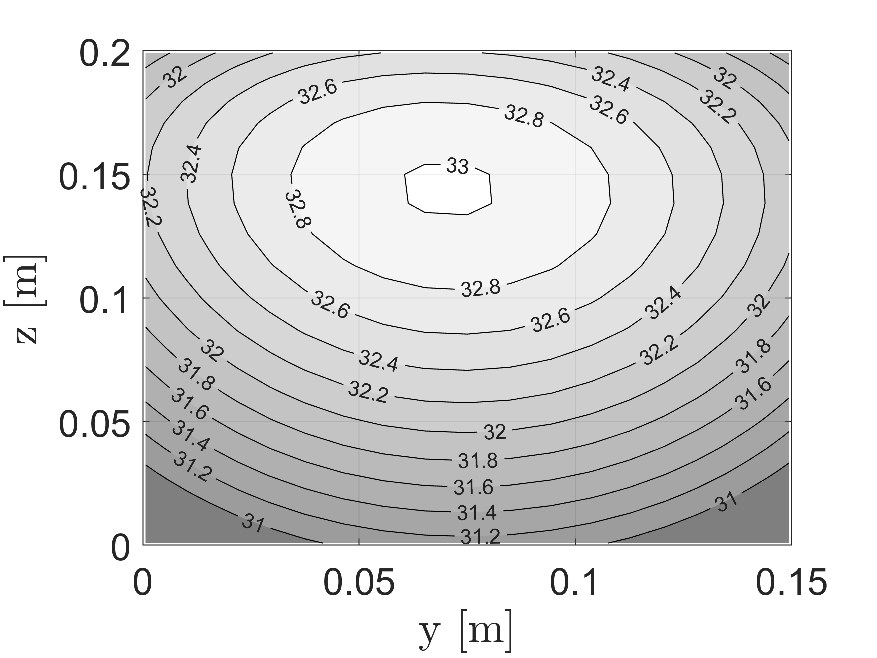}
                 \caption{Model: $t = 2500 $s. }
     \end{subfigure}
\hspace{1cm}
     \begin{subfigure}[t]{0.4\textwidth}
         \centering
         \includegraphics[width=\textwidth]{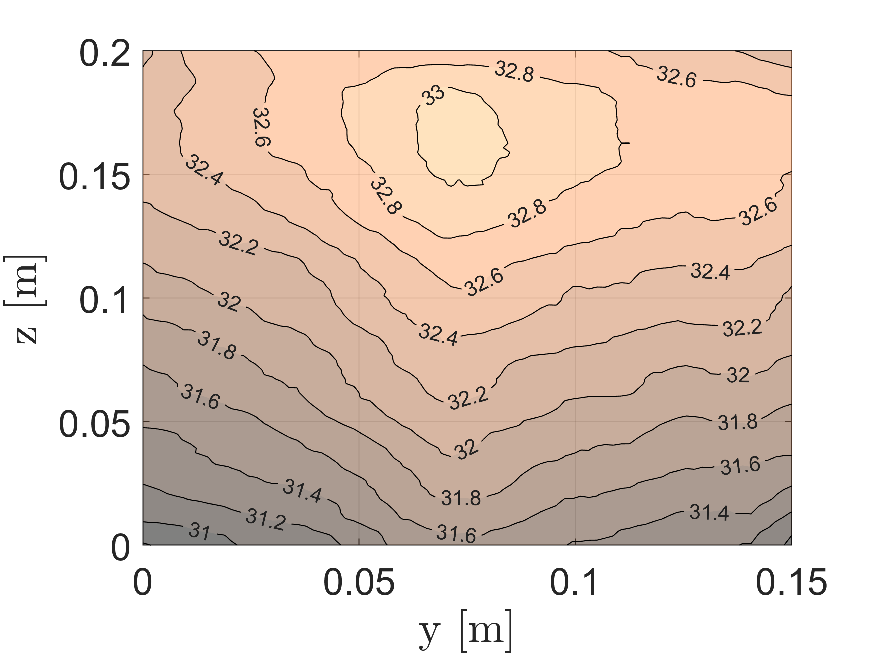}
         \caption{Data: $t = 2500 $s. }
     \end{subfigure}
        \caption{Temperature distributions across the plane of the pouch cell. Comparison between the model of \eqref{model_eqns} and the experimental data. Snapshots at $t$ = 100s, 500s, 1000s and 2500s are shown.  }
        \label{fig:snapshot}
\end{figure}

Figures \ref{fig:valid} and \ref{fig:snapshot} compares the model simulations against the experimental data\cite{lin2022multiscale} for cycling using the parameters from Table \ref{tab:params}. The parameters were estimated using a combination of fitting and a comparison to values from the literature, with the notation $R_0$ for the resistance of the uniform cell. Model agreement was observed in both the electrical and thermal response, as shown in Figures \ref{fig:valid_volt} and \ref{fig:valid_temp}. 

Figure \ref{fig:valid_temp} shows that the model was able to capture the evolution of the maximum, minimum, and average values of the cell temperature across the plane of the pouch cell during the cycling. A main discrepancy between the results of the model and the experimental data is that the results for the model are much smoother. The increased variability of the data is to be expected, as the sensing noise and the inherent randomness in the microstructure introduced by manufacturing limitations are not captured by Eqn. \eqref{model_eqns}.  However, the model is still able to capture the general trends of the data, with the time constants and steady-state values of the first-order responses seen in the temperature dynamics being captured.

Figure \ref{fig:snapshot} compares snapshots in the plane of the temperature distribution of the model and the experimental data \cite{lin2022multiscale}, where the snapshots were taken at $t =$ 100s, 500s, 1000s, and 2500s. Even though the underlying model of Eqn. \eqref{model_eqns} is simple, strong agreement was made with the experimental thermal data from Lin et al.,~2022\cite{lin2022multiscale}. In particular, the formation of a thermal hot spot as well as the general shape of the temperature contours were captured by the model, although, again, the model's diffusion dynamics resulted in smoother contours than those seen in the data.  \rd{We note that Lin et al.,~2022\cite{lin2022multiscale} presented a 3D model solved in COMSOL for the same data and a similar fit to that of Figure \ref{fig:snapshot} was achieved. Therefore, even though the model of  Eqn. \eqref{model_eqns} is arguably simpler than that of Lin et al.,~2022\cite{lin2022multiscale}  (see Section \ref{sec:NumSol} for details),  a similar level of model accuracy could be achieved.}

\begin{figure}
     \centering
     \begin{subfigure}[t]{0.32\textwidth}
         \centering
         \includegraphics[width=1\textwidth]{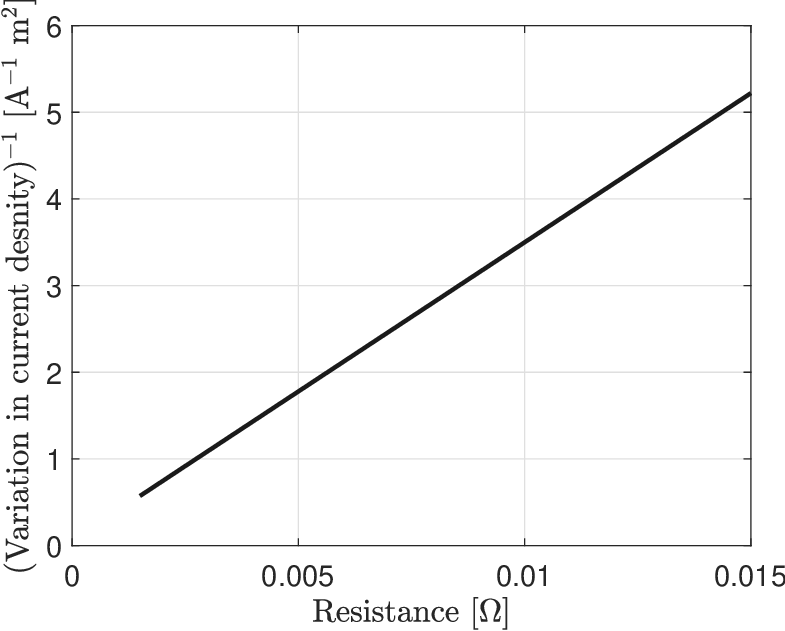}
       \caption{Current variation with resistance. }
   \label{fig:R0}
     \end{subfigure}
     \begin{subfigure}[t]{0.32\textwidth}
         \centering
         \includegraphics[width=1\textwidth]{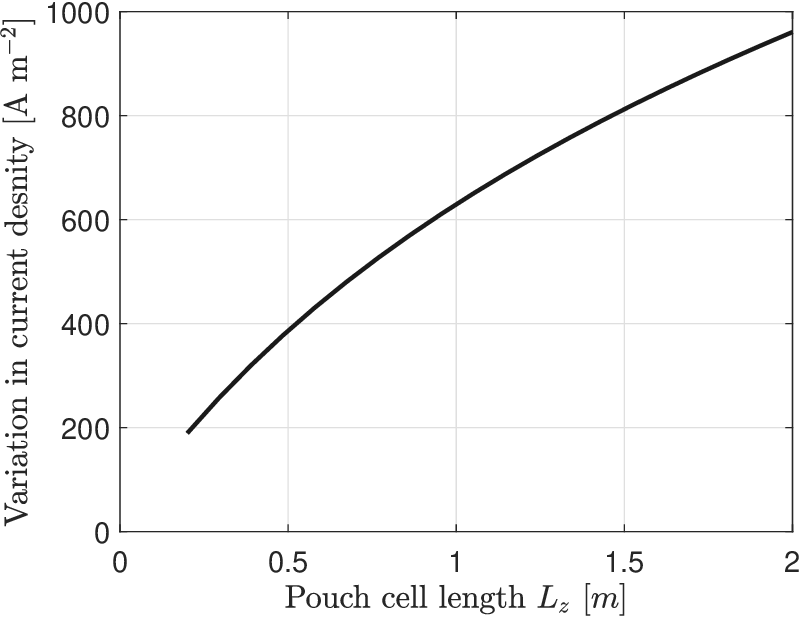}
                 \caption{Current variation with length.}
    \label{fig:Lz}
     \end{subfigure}   
 \begin{subfigure}[t]{0.32\textwidth}
         \centering
         \includegraphics[width=1\textwidth]{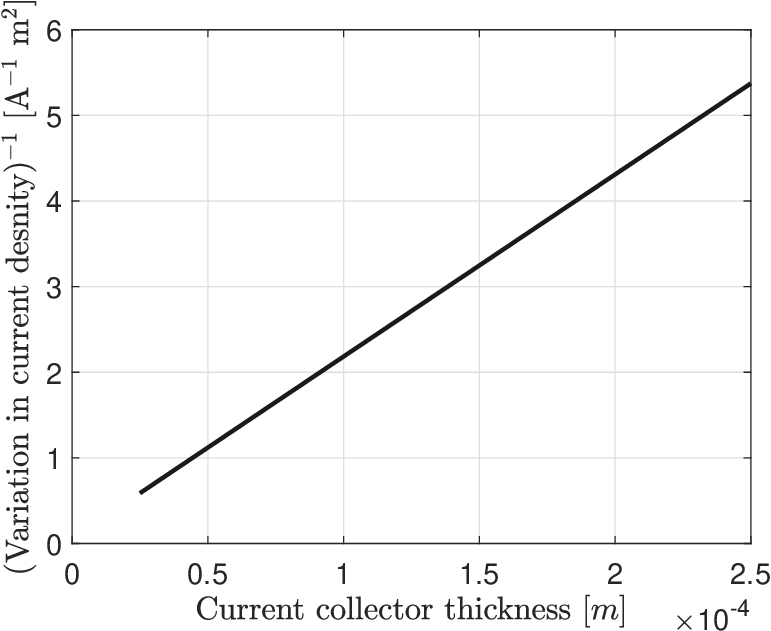}
                \caption{Current variation with current collector thickness. }
   \label{fig:Lcc}
     \end{subfigure}
     \hfill  
        \caption{Variation of the current distribution (difference between its maximum and minimum)  with pouch cell resistance,  length, and current collector thickness. }
        \label{fig:variations}
\end{figure}

Figure \ref{fig:variations} illustrates how the validated model can be used to infer the impact of design changes in the pouch cell on the degree of spatial heterogeneity across the plane. Defining the variation in the current density as the difference between the  maximum and minimum current density across the plane, Figure \ref{fig:R0} shows how this variation changes with resistance (as in, with $R(y,z) = R_0$),  Figure \ref{fig:Lz} shows how it changes with pouch cell length, and Figure \ref{fig:Lcc} shows how it changes with current collector thickness. The figure identifies some near linear relationships between these cell parameters and the variation in current density \rd{(Figure \ref{fig:variations} (b)), or its inverse (Figure \ref{fig:variations} (a) and (c))}, with explicit characterisations for these expressions obtainable from the analytical solution of Taheri et al.,~2014\cite{taheri2014theoretical}. As could be expected, the figure indicates that the variation of the current distribution across the pouch cell increases as: \textit{i)} the cell resistance decreases, \textit{ii)}  the aspect ratio increases, and  \textit{iii)}  the current collectors get thinner. Notably, Figure \ref{fig:R0} suggests that as the cell ages, and so as its resistance increases, the variation in current distribution across the cell is likely to decrease, as more current is encouraged to flow along the current collectors rather than through the thickness of the cell.

\section{In-the-plane electrode grading}\label{sec:grading}

The spatial distributions seen in the response of the pouch cell, for example in the temperature data of Figure \ref{fig:snapshot}, presents an opportunity to apply electrode grading to create a more uniform electrochemical response. In this section, an analytical solution for the resistance distribution across the plane of the pouch cell, $R(y,z)$, to achieve a uniform current distribution is described.


\subsection{Grading manufacturing}\label{sec:optimisation}
It is assumed that the electrode resistance $R(y,z)$ can be controlled locally in space. The reasons for focusing on the resistance distribution  are: \textit{i}) following Eqns. \eqref{eq1} and \eqref{eq2}, the electrode resistance $R(y,z)$ directly influences the current distribution across the pouch cell (and it is variations in the current which are targeted to be smoothed out by the optimally graded electrode), \textit{ii}) existing results on through-thickness electrode grading have demonstrated how this resistance grading can be achieved, e.g. by controlling the relative mass fraction of carbon black in LFP cathodes using spray printing techniques \cite{drummond2022modelling}. \rd{Achieving in-plane grading of electrodes using conventional techniques can be challenging. However, innovative methods such as spray printing show promise for creating a gradient across the thickness of electrodes. This technique can also effectively apply grading across the plane of the electrode. While spray printing has potential, it is currently limited to small-scale applications. In contrast, traditional methods like slurry casting can be creatively adapted to achieve partial in-plane grading\cite{cheng2020combining,cheng2022extending}. In future research, we will present experimental data on in-plane graded LFP electrodes fabricated using both spray printing and modified slurry casting techniques.}


\subsection{Resistance distribution for homogeneous current}
The problem of distributing the resistance (achieved by controlling the weight fraction of active carbon in the electrode microstructure) to achieve homogeneous current distribution across the pouch cell is considered. This targeted flat current distribution is equal to the average current density, with
\begin{align}\label{eqn:flat_current}
    i_x(y,z,t) = \frac{I_{\text{app}}(t)}{A_\text{cell}}.
\end{align}

Eqn. \eqref{model_eqns} implies that, under isothermal conditions of the open-circuit voltage, if a uniform current distribution is achieved at $t = 0$, then $v_1$, $v_2$ and the SoC, will evolve uniformly in space as well. The implication is that if the current distribution is uniform at $t = 0$, it will remain uniform during the simulation, following the model equations of \eqref{eq1} and \eqref{eq2}, which may not hold exactly in practice as local current spikes might emerge caused by the variability in the local electrode composition. However, this can be used to simplify the graded electrode design optimisation problem, \rd{and smooth out current and temperature gradients across the pouch cells}. Specifically, it converts a dynamic optimisation problem into a static one, in the sense that it changes the optimisation from one of determining how the resistance should be distributed to achieve a uniform current distribution during the whole charge to one that only has to consider the distribution at the start of the charge. A consequence of this problem formulation is that it allows an analytical solution to the graded electrode design optimisation to be derived (see Eqn \eqref{eqn:sol}).

\subsubsection{Problem formulation}
The desired uniform current distribution of Eqn. \eqref{eqn:flat_current} is substituted into Eqns \eqref{eq1} and \eqref{eq2} to give, after simplifying,
\begin{subequations}\label{eqn:mod_R}\begin{align}
\Delta R -\frac{1}{A_\text{cell}\sigma_\text{eff}(L_\text{cn}+L_\text{cp})} = 0, \ \  \text{on} \ \   \Omega ,
\end{align}
which is a non-homogeneous Laplace equation in the spatial variable $R(y,z)$. Using the same approach, the  boundary conditions for the voltages from Eqns. \eqref{bveq2} and \eqref{bveq4} can be translated into boundary conditions for the resistance to achieve a uniform current density, as in 
\begin{align}
				\frac{\partial R}{\partial z} \Big|_{  z = L_z   }    	& = \frac{ 
1 }{ \sigma_{\scriptscriptstyle \text{ck}}   	A_{\scriptscriptstyle \text{tab},k}}   ,\ \  \text{on} \ \  \partial \Omega_{\scriptscriptstyle \text{tab,k}}  ,\,k \in\{n,p\}, 
\\
n_{\perp} \cdot \nabla R	& =    0   ,\ \  \text{on} \ \  \partial \Omega_{\scriptscriptstyle \text{ext}}.
\end{align}\end{subequations}

\subsubsection{Solution}
The solution to  Eqn. \eqref{eqn:mod_R} is
\begin{subequations}\label{eqn:sol}\begin{align}
R(y,z) = R_\text{quad}(z)+ R_\text{hyper}(y,z)
\end{align}
composed of a quadratic part, $R_\text{quad}(y,z)$, and a hyperbolic part,  $R_\text{hyper}(y,z)$, satisfying
\begin{align}
R_\text{quad}(z) & = d_0+d_1z+d_2z^2,
\\
R_\text{hyper}(y,z)  & = \sum_{n = 0}^\infty c_{n}\cos\left(\frac{n\pi y}{L_y}\right)\cosh\left(\frac{n\pi z}{L_y}\right).\label{hyper_r}
\end{align}
The coefficients of this solution are 
\begin{align}
    d_1  & = 0, \\
    d_2 &=  \frac{1}{2\; A_\text{cell}\;\sigma_\text{eff}\;(L_\text{cn}+L_\text{cp})}, \\
    c_{n}  & = \frac{2}{n \pi\sinh\left(\frac{n\pi L_z}{L_y}\right)} \int^{L_y}_0h(y)\cos\left(\frac{n\pi y}{L_y}\right)~dy,
\end{align}
with
\begin{align}
h(y) 
= \begin{cases}  -\frac{L_z}{A_\text{cell}\sigma_\text{eff}(L_\text{cn}+L_\text{cp})} & y \in \Omega_\text{ext},~ z = L_z,
\\  \frac{1 }{ \sigma_{\scriptscriptstyle \text{c1}}   	A_{\scriptscriptstyle \text{tab},1}} -\frac{L_z}{A_\text{cell}\sigma_\text{eff}(L_\text{cn}+L_\text{cp})} & y \in \Omega_{\scriptscriptstyle \text{tab,1}},~ z = L_z,\\
 \frac{1 }{ \sigma_{\scriptscriptstyle \text{c2}}   	A_{\scriptscriptstyle \text{tab},2}} -\frac{L_z}{A_\text{cell}\sigma_\text{eff}(L_\text{cn}+L_\text{cp})} & y \in \Omega_{\scriptscriptstyle \text{tab,2}},~ z = L_z, \end{cases}
\end{align} \end{subequations}
and the constant $d_0$ being a free variable  setting the average value of the resistance. This resistance distribution defines the graded electrodes of the following analysis. \rd{To evaluate Eqn. (7c), the infinite sum has to be truncated after a finite number of terms. It was found that a truncation of the order of 20 terms was sufficient to achieve a stable and accurate solution. }

The solution of Eqn \eqref{eqn:sol} shows that the resistance should roughly decrease  away from the tabs in order to flatten the current distribution. The physical meaning of  decreasing the resistance away from the tabs is to encourage current to flow down the length of the current collectors, instead of going directly through the electrodes as it enters one of the tabs and leaves the other. In the particular setup of Figure \ref{fig:pouch_model} and Table \ref{tab:params},  where the cell has a relatively high aspect ratio, the variation of the resistance is much larger in the $z$-direction rather than the $y$-direction, but, following Eqn. \eqref{eqn:sol}, this may change for wider cell formats or for different tab configurations. 

\begin{figure}
     \centering
     \begin{subfigure}[t]{0.35\textwidth}
         \centering
         \includegraphics[width=\textwidth]{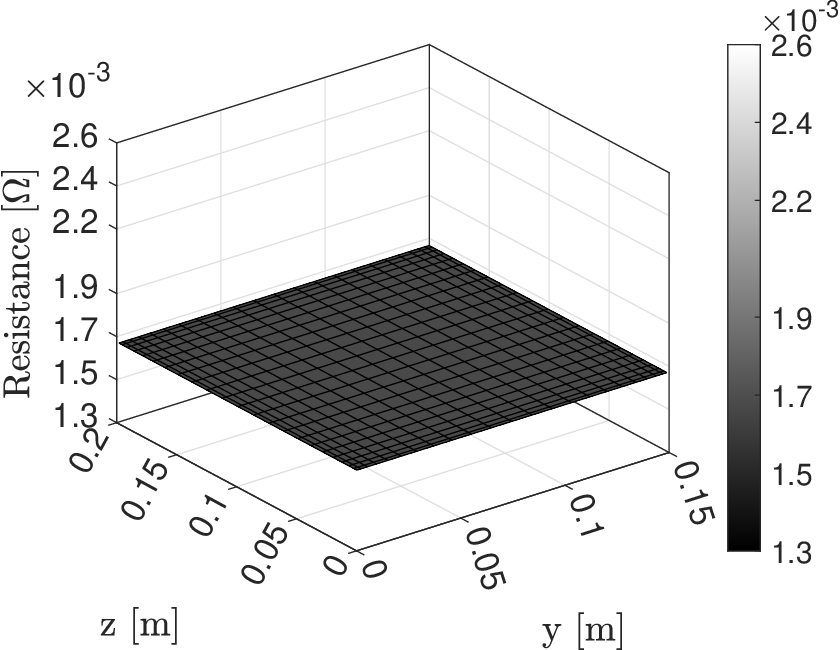}
                \caption{Resistance of the uniform cell.}
         \label{fig:res_unif}
     \end{subfigure}     \hspace{1cm}
     \begin{subfigure}[t]{0.35\textwidth}
         \centering
         \includegraphics[width=\textwidth]{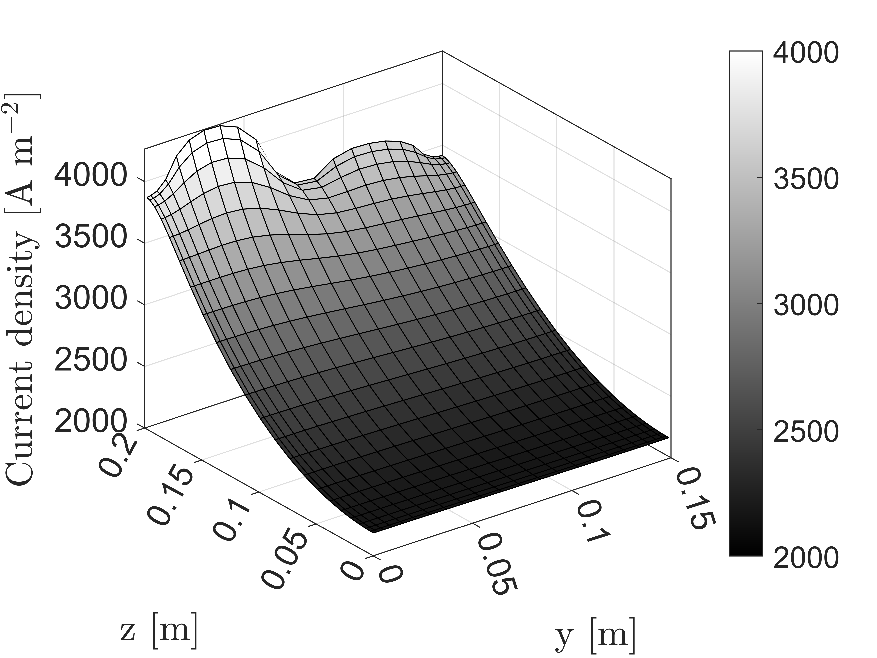}
                 \caption{Current density of the uniform cell. }
         \label{fig:curr_unif}
     \end{subfigure}    \\    
 \begin{subfigure}[t]{0.35\textwidth}
         \centering
         \includegraphics[width=\textwidth]{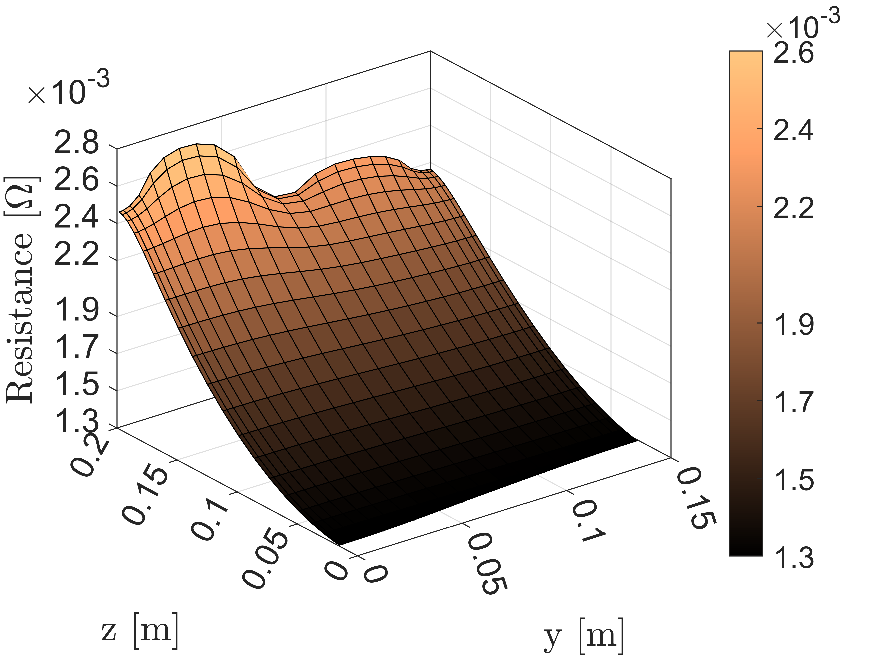}
                \caption{Resistance of the graded cell.}
         \label{fig:res_grad}
     \end{subfigure}
   \hspace{1cm}
     \begin{subfigure}[t]{0.35\textwidth}
         \centering
         \includegraphics[width=\textwidth]{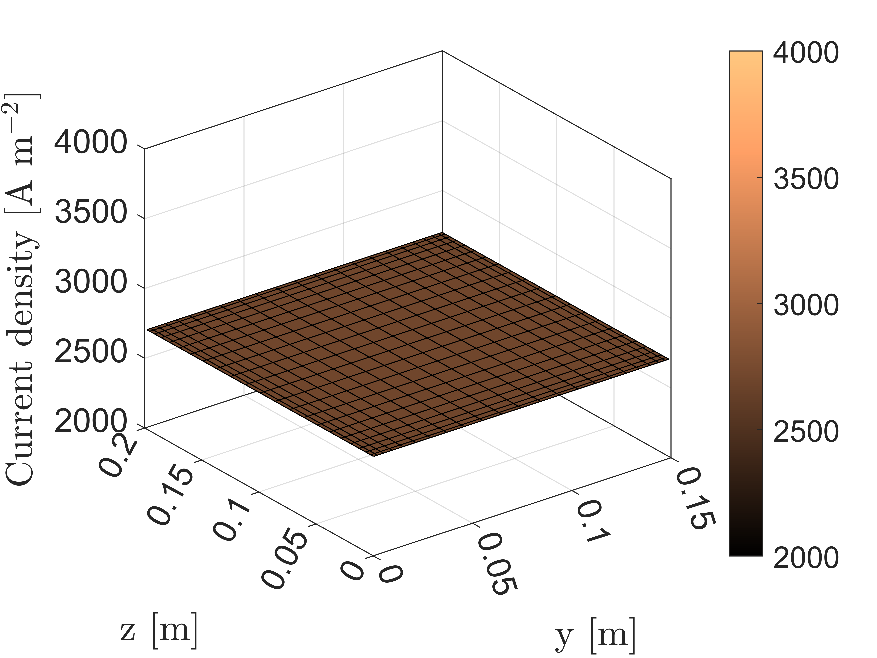}
       \caption{Current density of the graded cell.  }
   \label{fig:curr_grad}
     \end{subfigure}
        \caption{Distributions in resistance and current density of the graded and uniform cells after a 4C charge. The uniform cell has a constant resistance, but a spatially varying current distribution. The graded cell has a spatially varying resistance, but a constant current distribution.   }
        \label{fig:curr_res}
\end{figure}

\subsection{Results: Flat current distribution across pouch cells}

Figure \ref{fig:curr_res} compares the current and resistance distributions achieved with both uniform  and graded pouch cells (with the resistance of the graded cells following the solution derived in Eqn. \ref{eqn:sol}). For this analysis, both pouch cells had equal average resistances of 1.5 $\times 10^{-3}$  $\Omega$ to ensure a fair comparison. Significant variation was seen in the current density distribution of the uniform cell (with a maximum value of 3925 A m$^{-2}$ at the tabs and a minimum value of 2138 A m$^{-2}$ at the opposite end where $z = 0$). By contrast, the graded cell achieved a constant current density of 2667 A m$^{-2}$ across the plane, as predicted. 

\rd{Regarding whether the optimised resistance distributions of Figure \ref{fig:curr_res} could be realised in practice, consider  the maximum change in resistance that could be achieved by changing the local weight fraction of carbon black in the LFP cathodes. The characterisation of electronic conductivity of LFP cathodes in terms of the weight fraction of carbon black, $w_\text{cb}$, from Drummond et al.,~\cite{drummond2022modelling} is assumed. The resistance of the LFP cathode, $R_{\text{LFP}}$, is  approximated\cite{drummond2022modelling} by 
\begin{align}
    R_{\text{LFP}} = \frac{\ell}{n_\text{layers} \times  A_\text{cell} \times \sigma_\infty \times  \  {w_\text{cb}}^b }
\end{align}
where $\sigma_\infty = 4.01$ S m$^{-1}$, $b = 1.7$, and the electrode thickness is $\ell = 100$ $\mu$m. Define $\overline{R}_{\text{LFP}}$ as the maximum resistance in Figure \ref{fig:curr_res} (with a weight fraction of carbon black of $\overline{w}_\text{cb}$) and $\underline{R}_{\text{LFP}}$ as the minimum resistance (with a weight fraction of carbon black of $\underline{w}_\text{cb}$). Note that $\underline{w}_\text{cb} > \overline{w}_\text{cb}$ since carbon black increases the conductivity of the LFP electrode.  The difference between the maximum and minimum resistances  of the graded design is $\overline{R}_{\text{LFP}}-\underline{R}_{\text{LFP}}\approx 1.2 \times 10^{-3}\,\Omega$. Therefore, the differences between the two weight fractions satisfies
\begin{align}
\frac{1}{ {\overline{w}_\text{cb}}^b}-\frac{1}{ {\underline{w}_\text{cb}}^b}
=
    \frac{n_\text{layers} \times  A_\text{cell} \times \sigma_\infty\times \left(\overline{R}_{\text{LFP}}-\underline{R}_{\text{LFP}}\right)}{\ell }. \label{cbd_res}
\end{align}
If the weight fraction of carbon black in the LFP cathode at the point with minimum resistance is $\underline{w}_\text{cb} = 0.06$, then Eqn.~\ref{cbd_res} implies that the weight fraction at the point with  maximum resistance should be $\overline{w}_\text{cb} = 0.0471$.  This is a modest difference and readily achievable in practice. Subtle changes in the electrode microstructure, for example in the distribution of carbon black in LFP cathodes, can therefore significantly influence how heterogeneities emerge across pouch cells.}



\subsection{Results: Dynamic response of graded pouch cells}\label{sec: improvement}

\begin{figure}
     \centering
     \begin{subfigure}[t]{0.36\textwidth}
         \centering
         \includegraphics[width=0.95\textwidth]{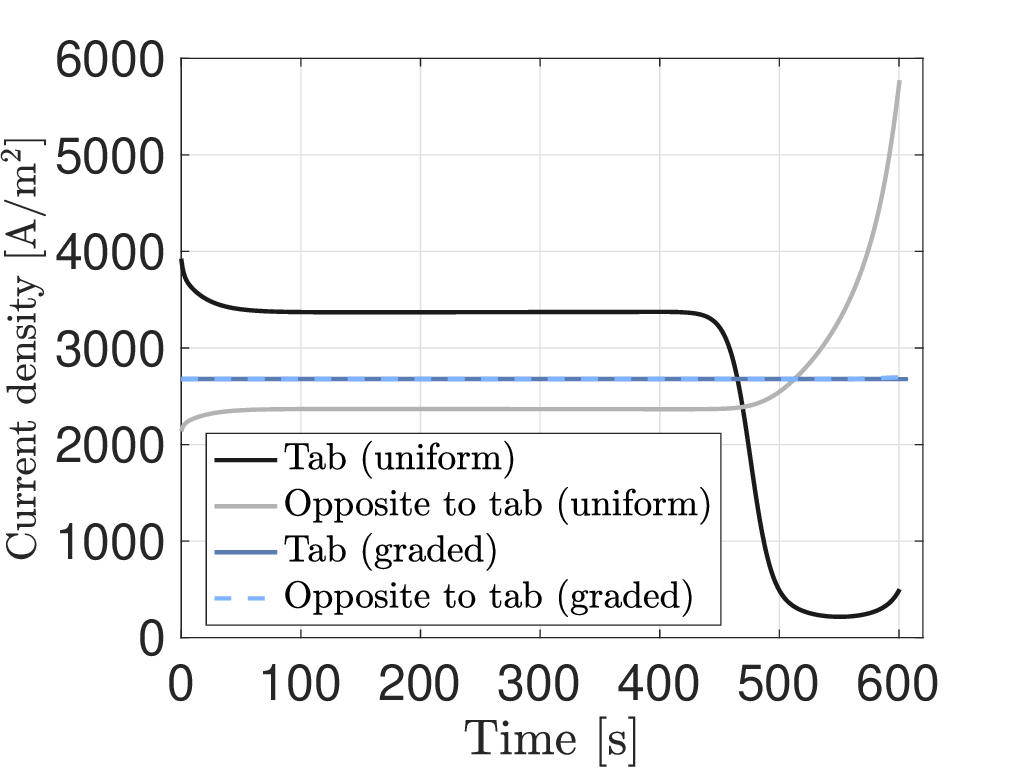}
       \caption{Current density. }
   \label{fig:comp_curr}
     \end{subfigure}
\hspace{1cm}
     \begin{subfigure}[t]{0.36\textwidth}
         \centering
         \includegraphics[width=0.95\textwidth]{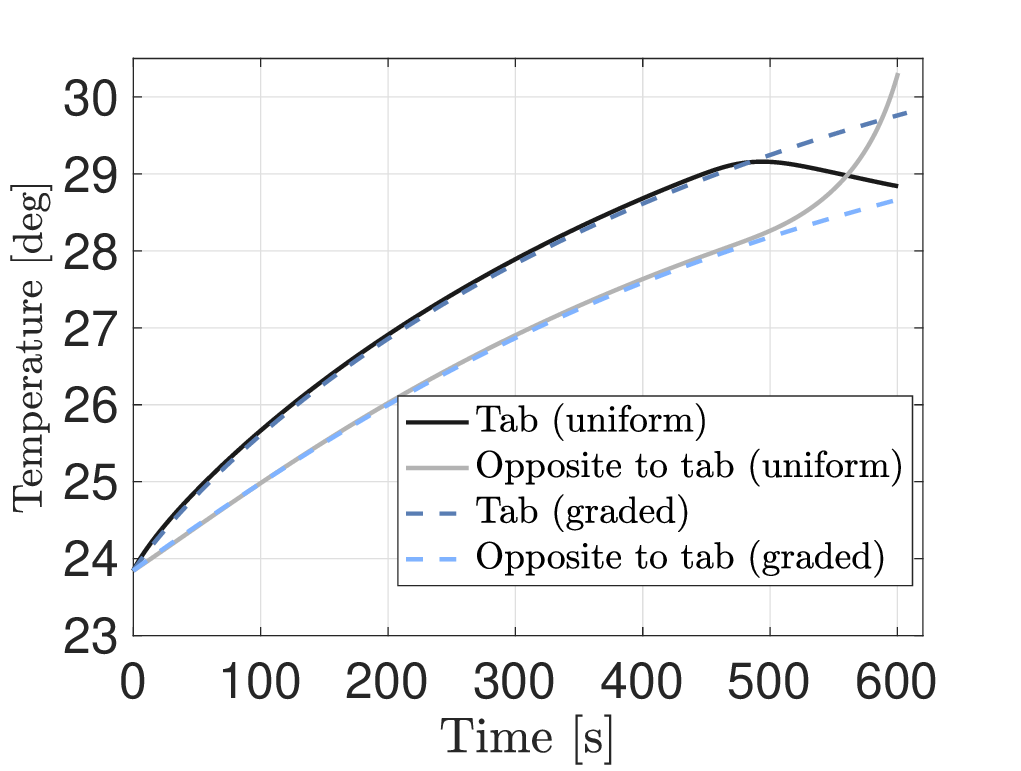}
                 \caption{Temperature.}
    \label{fig:comp_temp}
     \end{subfigure}    
\\
 \begin{subfigure}[t]{0.36\textwidth}
         \centering
         \includegraphics[width=0.95\textwidth]{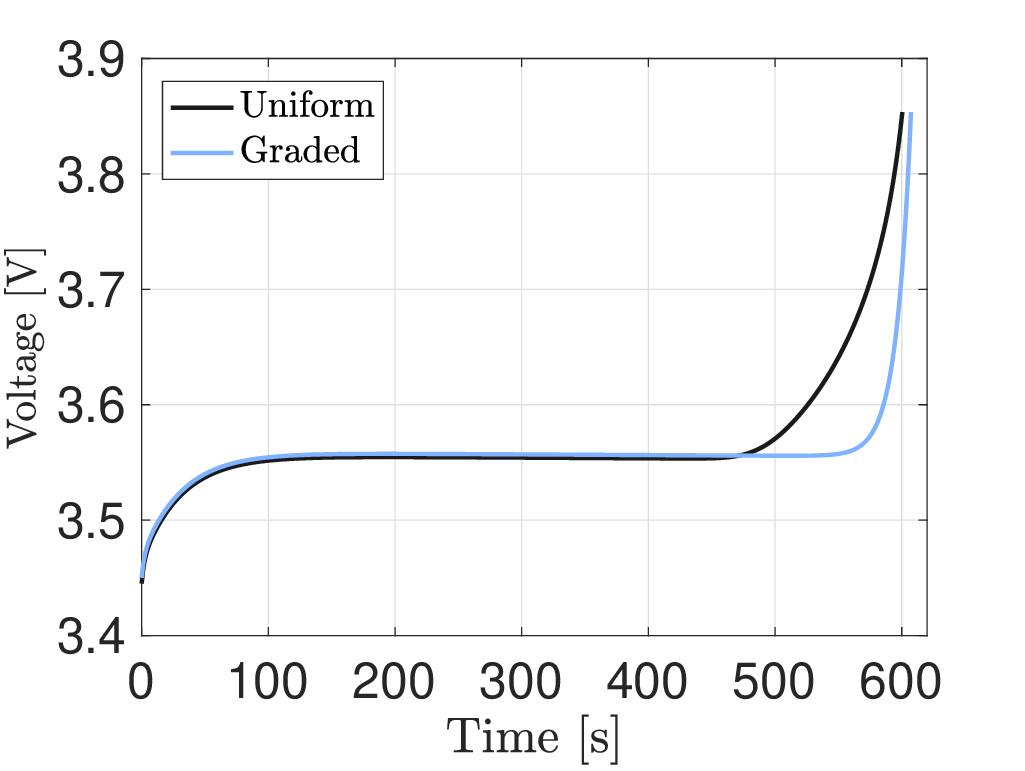}
                \caption{Voltage}
   \label{fig:comp_volt}
     \end{subfigure}
\hspace{1cm}
  \begin{subfigure}[t]{0.36\textwidth}
         \centering
         \includegraphics[width=0.95\textwidth]{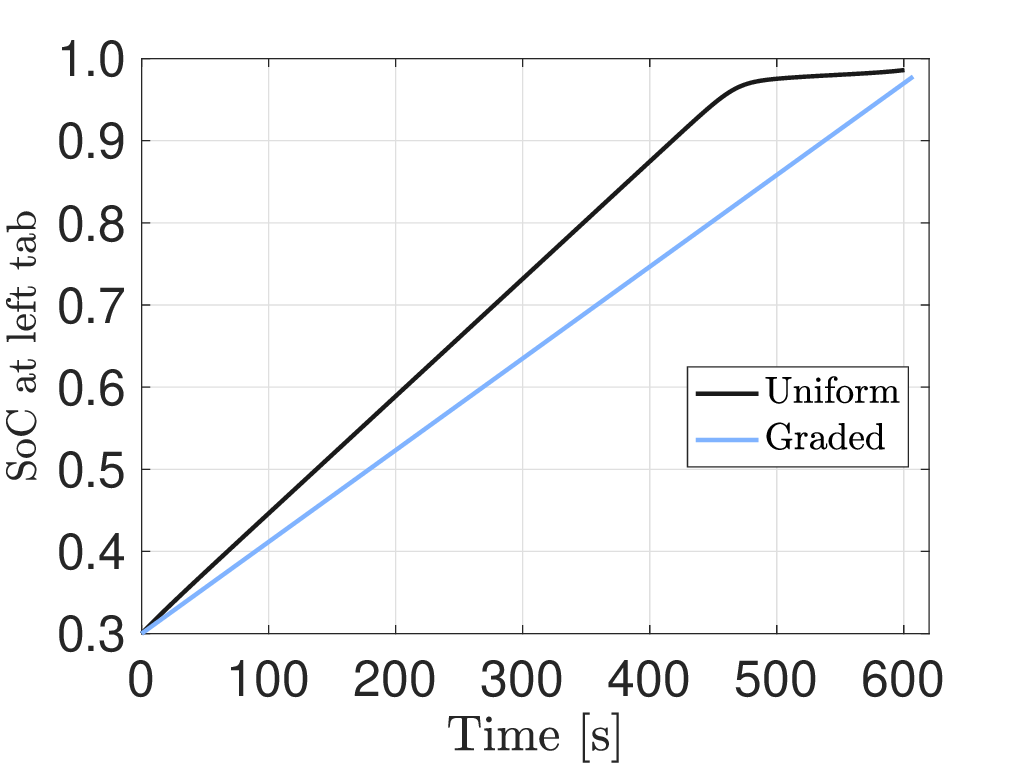}
                \caption{State-of-charge. }
   \label{fig:comp_soc}
     \end{subfigure}    
        \caption{Spatial-heterogeneity of the model's variables, including state-of-charge, temperature, voltage and current density during a 4C charge. The difference between these variables at the tabs and opposite the tab is shown. }
        \label{fig:comp}
\end{figure}

Figure \ref{fig:comp} compares the optimised graded electrode cell following Eqn \eqref{eqn:sol} and the uniform cell for a full 4C charge at $I_{\text{app}}$ = 80A.  
Figure \ref{fig:comp} provides justification for the design choice of Section \ref{sec:optimisation}, which optimises the spatial resistance of the pouch cells to achieve a uniform initial current distribution in space, as Figure \ref{fig:comp_curr} implies that this uniformity in the current was retained during the whole charge. By contrast, the current distribution in the uniform cell is more variable. Initially, large current gradients exist across the plane of the uniform cell, as shown in Figure \ref{fig:comp_curr}, with the maximum current density 3925 A/m$^2$ at the tabs and the minimum current 2138 A/m$^2$ at the opposite end. The non-uniform current distribution then causes the voltage to vary, following Eqns. \eqref{eq4}-\eqref{eq6}. Simultaneously, there is a negative feedback effect acting on the voltages from Eqns. \eqref{eq1} and \eqref{eq2} that smooths this variation. The interactions between these two modes causes the spatial distribution of the current to evolve in both space and time. In particular, Figure \ref{fig:comp_curr} shows that the current does not change significantly during 100s $ \leq t \leq $450s when the OCV curve is flat (since the smoothing effect of $U$ in Eqn. \eqref{eq1} is not active). It is this smoothing effect that causes the region near the tabs to charge first, but then, when it is fully charged, the majority of the current then flows through the bottom half of the cell.


\begin{figure}
     \centering
     \begin{subfigure}[t]{0.35\textwidth}
         \centering
         \includegraphics[width=\textwidth]{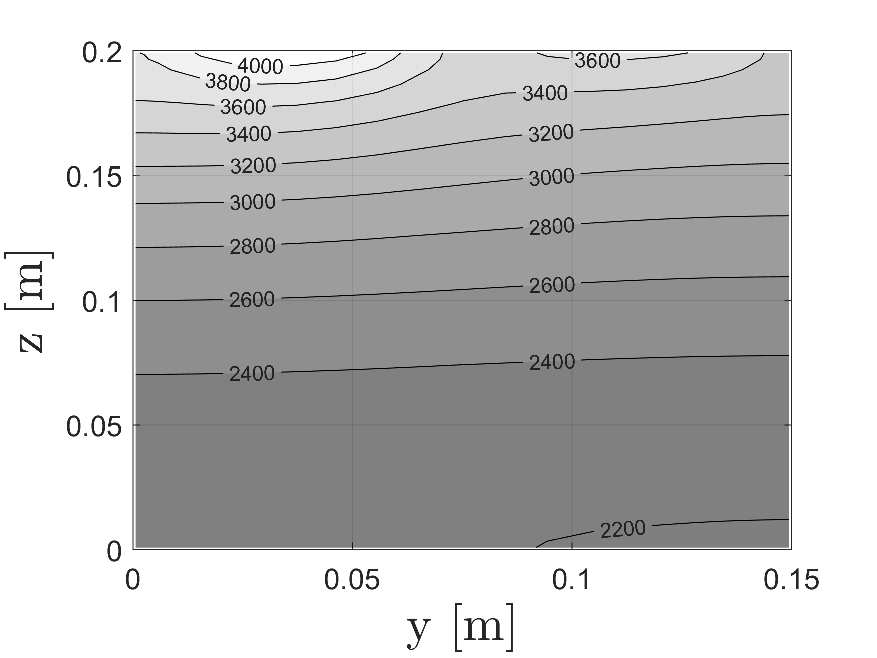}
                \caption{Current density: $t = 1 $s. }
     \end{subfigure}     \hspace{1cm}
     \begin{subfigure}[t]{0.35\textwidth}
         \centering
         \includegraphics[width=\textwidth]{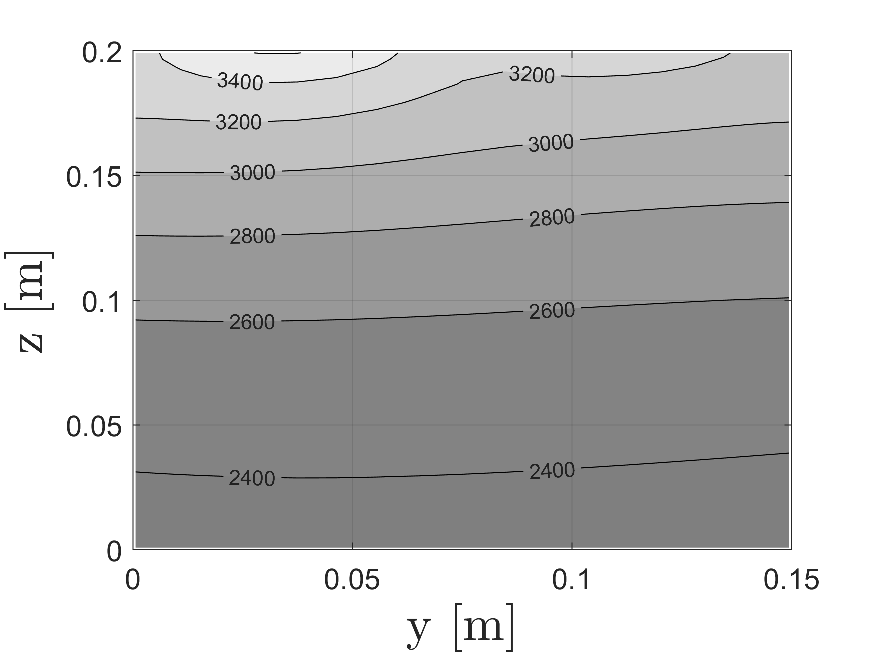}
                \caption{Current density: $t = 200 $s. }
     \end{subfigure}     \hfill
 \begin{subfigure}[t]{0.35\textwidth}
         \centering
         \includegraphics[width=\textwidth]{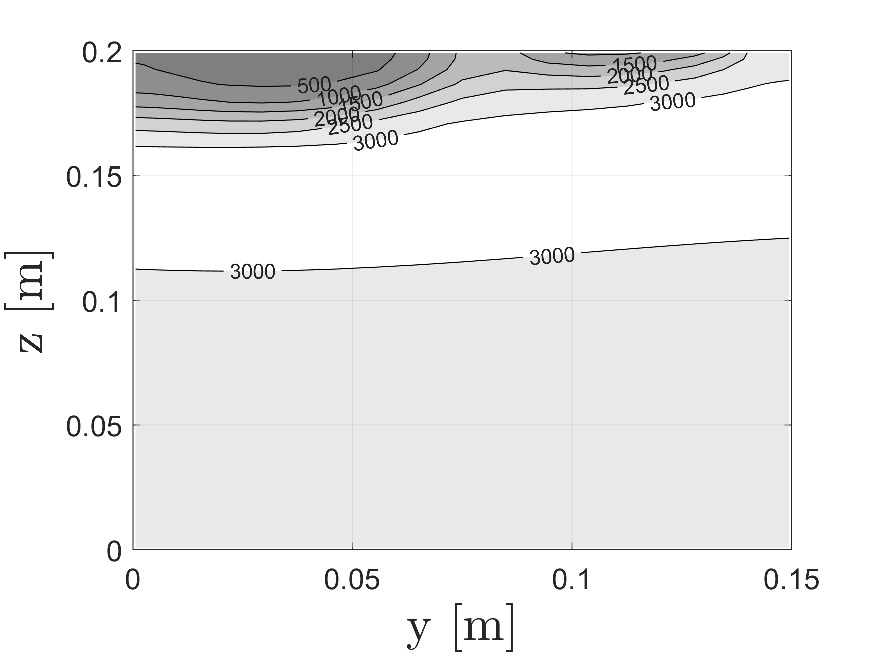}
                \caption{Current density: $t = 500 $s. }
     \end{subfigure}
     \hspace{1cm} 
     \begin{subfigure}[t]{0.35\textwidth}
         \centering
         \includegraphics[width=\textwidth]{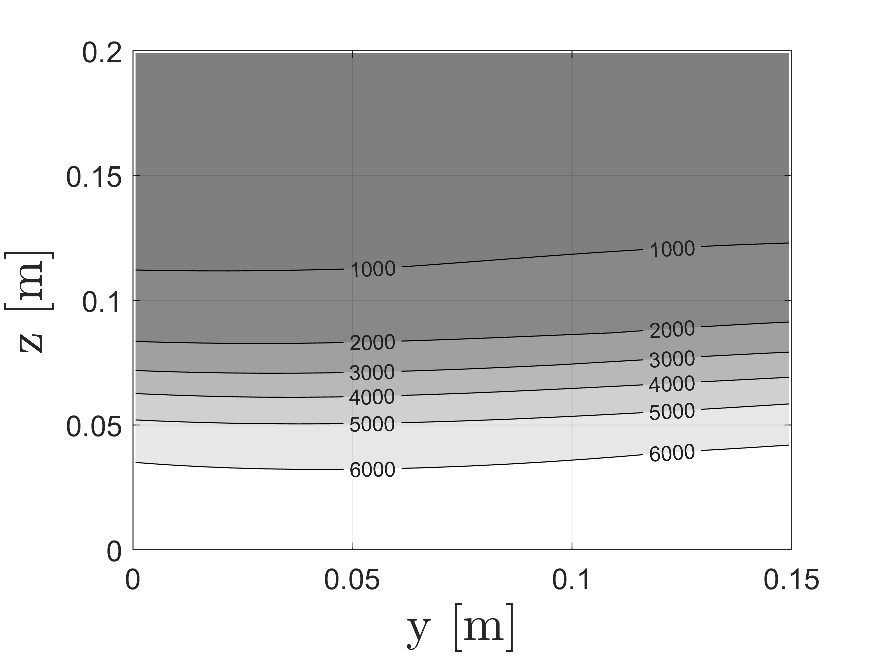}
                \caption{Current density: $t = 600 $s. }
     \end{subfigure}
         \caption{Current density distributions in the plane of the uniform pouch cell using the model of Eqn. \eqref{model_eqns} during a constant current charge at 4C. Contour lines indicate current density values in A m$^{-2}$ going through the electrode sandwich. The location of maximum current travels from the tabs to opposite them during the charge. }
        \label{fig:snapshots_curr}
\end{figure}

Figure \ref{fig:snapshots_curr} shows snapshots of the current distribution across the current collectors of the pouch cell with uniform electrodes at $t$ = 1s, 200s, 500s and 600s during the 4C charge. At the start of the simulation, the peak current was  at the tabs, with this peak moving down the cell during the charge. It is noted that the maximum variation in the current distribution in the plane was recorded near the end of the charge, which is also the time when lithium-plating is most likely to occur. 

 Whilst Figure \ref{fig:comp_curr} implies that there would be large differences between the current distributions of the uniform and graded cells, Figures \ref{fig:comp_temp} and \ref{fig:comp_soc} indicate that the impact of these differences on the variable \emph{internal} to the current collectors would actually be quite limited. Specifically, Figure \ref{fig:comp_temp} compares the temperatures, Figure \ref{fig:comp_volt} the voltages, and Figure \ref{fig:comp_soc} the state-of-charges. Although there are some differences between the uniform and graded electrodes, it is relatively limited compared to the large variations seen in the current distributions. In particular, even though the profiles of the two voltage curves in Figure \ref{fig:comp_volt} are different, they reach the cut-off voltage of 3.85 V at approximately the same time (607s for the graded cell and 600s for the uniform one). As such, the capacitance improvement of the graded cell over the uniform one is only 1.2\%. These limited gains in the capacitance are a result of the large thermal and electrical conductivities of the current collectors (see Table \ref{tab:params}), which smooth out the variations in the electrodes responses across the plane.  Similarly, for the current, Appendix Figure~\ref{fig:GradedvsUniformcycled} compares the modelled 4C cycling results for the graded and uniform cells.  The current density of the graded cell is close to $I(t)/A_\text{cell}$, while the uniform cell has large variations both in time and across the plane. By contrast, the maximum temperature is slightly reduced by $0.3\,^{\circ}\mathrm{C}$ for the graded cell.
 
When only variables defined internally to the current collectors (such as temperature and voltage) are used to assess the value of graded pouch cells, their benefits may be lost. This observation agrees with the conclusions of Hosseinzadeh et al.,~2018\cite{hosseinzadeh2018impact} that porosity distribution  through the electrode thickness has the potential to produce superior battery performance instead of when the porosity is varied along the electrode height. However, in this paper, it is proposed that to unlock the benefits of graded pouch cells, a performance metric that focuses on \emph{through-thickness effects} should instead be considered. This observation motivates the analysis of the following section on the potential benefits of graded electrodes to mitigate lithium plating.

\section{Results: Plating reduction with graded electrodes}\label{sec:plating}

The potential of graded electrodes to mitigate lithium plating in large format cells is explored in this section. As discussed in relation to Figures \ref{fig:comp_temp}-\ref{fig:comp_soc} and in the results of Hosseinzadeh et al.,~2018\cite{hosseinzadeh2018impact}, it was observed that the high conductivity of the current collectors limits the apparent value of electrode grading in-the-plane when only variables found ``\textit{internally}'' to the current collectors (with  \textit{internal} variables being understood as those, such as voltage and temperature, that are smoothed out when flowing in the plane of the current collectors) are used to assess performance, as opposed to variables flowing through the electrodes (such as the current distributions of Figure \ref{fig:snapshots_curr} and \ref{fig:comp_curr}). However, as lithium plating is a localised degradation phenomenon dependent upon the charging current and electrochemical state of the electrode at a given point,  it is shown in this section that plating may  be effectively mitigated by in-the-plane electrode grading. The focus towards mitigating lithium plating is also motivated by experimental studies \cite{yang2019asymmetric} which show significant spatial variations in plating across graphite anodes. Controlling the distribution of active carbon in the electrode microstructure to  mitigate the spatial distribution in plating is proposed as a means to reduce the accelerated degradation of large format pouch cells in high-rate applications.

A mathematical modelling approach was used to compare the extent of lithium plating between the graded and uniform cells. However, modelling lithium plating is known to be challenging, with existing models typically being additions to complex electrochemical models, with example models discussed in \cite{o2022lithium, reniers2019review}. By contrast, in this paper, a simpler, control-orientated model adapted from \cite{couto2021faster} was used, with this model based off of the earlier results of \cite{romagnoli2019feedback}. For the state-constrained control problem considered in \cite{couto2021faster}, the  following condition for lithium plating was developed, with the anode regarded as experiencing plating if the following inequality is satisfied 
\begin{align}\label{eqn:plating}
a_\text{lp}\log(b_\text{lp}\text{SoC}(y,z,t))+c_\text{lp}+d_\text{lp}i_x(y,z,t)  \geq 0.
\end{align}
 The plating inequality\cite{couto2021faster} of Eqn. \eqref{eqn:plating} was determined by fitting a curve when the anode over-potential in a DFN model went negative.  Here, $a_\text{lp} = 1.74$, $b_\text{lp} = 9.32$, $c_\text{lp} = -4.46$, following \cite{couto2021faster}, and $d_\text{lp} =0.0055$.

\begin{figure}
     \centering
     \begin{subfigure}[t]{0.48\textwidth}
         \centering
         \includegraphics[width=1\textwidth]{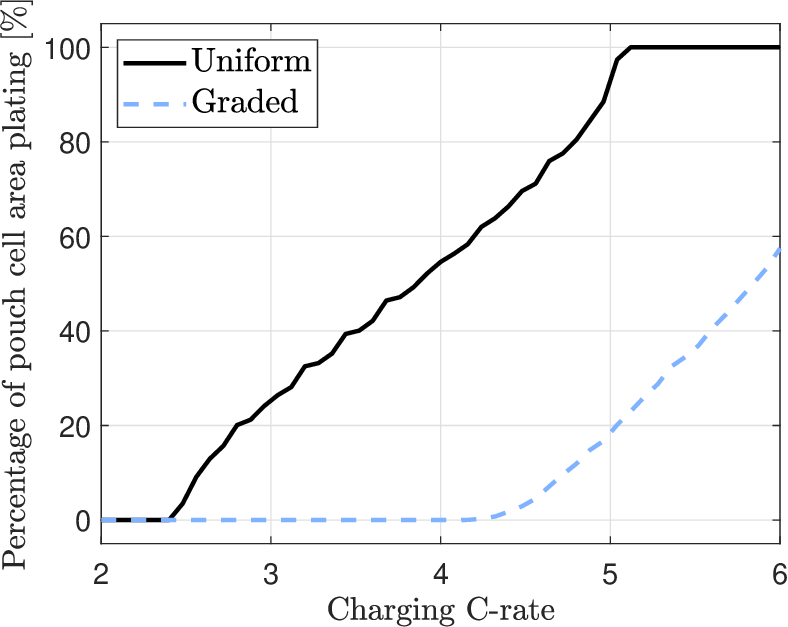}
       \caption{C-rate dependency of plating. }
   \label{fig:c_rate}
     \end{subfigure}
     \begin{subfigure}[t]{0.48\textwidth}
         \centering
         \includegraphics[width=1\textwidth]{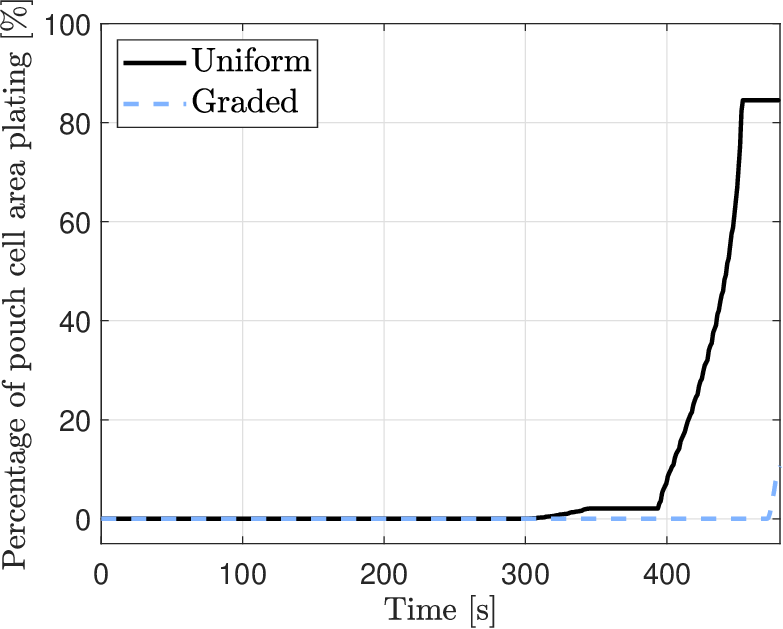}
                 \caption{Plating onset during 5C charging.}
    \label{fig:plating_time}
     \end{subfigure}   
     \hfill  
        \caption{Comparing the onset of lithium-plating between the uniform and graded pouch cells. Figure \ref{fig:c_rate} shows the C-rate dependency of the plating and Figure \ref{fig:plating_time} compares the progression of plating during 5C charging. 
 }
        \label{fig:plating}
\end{figure}

 \begin{figure}[h]
     \centering
\includegraphics[width=0.5\textwidth]{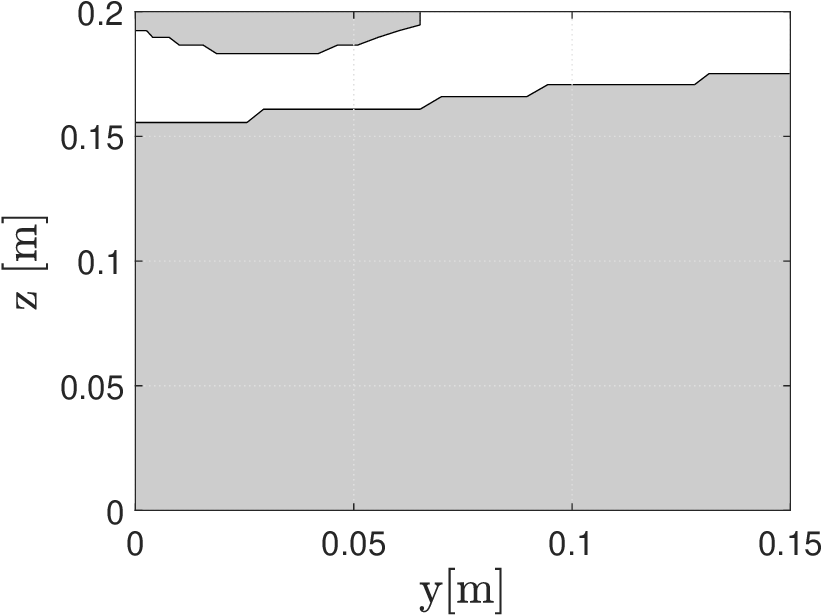}
                \caption{Distribution of plating across the plane at the end of a 5C charge of the uniform cell. The grey region has experienced plating and the white region is free of plating. }
   \label{fig:plating_space}
 \end{figure}
 
Figure  \ref{fig:plating} compares the onset of lithium-ion plating for both the uniform and graded pouch cells using Eqn. \eqref{eqn:plating}. The purpose of this figure is to illustrate the spatial, temporal and C-rate dependence of plating, and how modelling can capture the heterogeneity of pouch cell degradation. \rd{For this simulation, the temperature dependency of the onset of lithium plating was neglected as it was assumed that the temperature of the cell was approximately constant. This assumption could be relaxed using more complex plating models than Eqn. \eqref{eqn:plating}, such as those discussed in Reniers et al.,~\cite{reniers2019review}, as the likelihood of plating is known to increase as the cell temperature decreases, especially for sub zero temperatures\cite{zinth2014lithium}}. Figure \ref{fig:c_rate} shows the areal percentage of the pouch cell experiencing plating during charging at different C-rates. The uniform pouch cell sees a gradual increase in the area of  the pouch cell experiencing plating when the C-rate is increased beyond 2.4C, with 100\% of the pouch cell experiencing plating at 5.1C. By contrast, the graded pouch cell experiences no plating until 4.3C. The differences between these two curves is due to variation in the current distribution of the uniform cell; this variation leads to higher current spikes and so a greater likelihood of plating. By contrast, the model predicts that the flatter current distribution of the graded pouch cell can delay the onset of plating until  a critical point is reached, and, at that point, the flat distributions cause the whole plane to plate. \rd{The predicted critical predicted C-rates for lithium plating are broadly consistent with the experimental literature; for example, Burns et al.,~\cite{burns2015situ} detected plating in 220~mAh Li[Ni$_{1/3}$Mn$_{1/3}$Co$_{1/3}$]O$_2$/graphite 402035-size pouch cells at approximately 1C and 30$^\circ$ C}. The significance of Figure \ref{fig:c_rate} is that it suggests grading pouch cells could increase the maximum charging C-rate for which plating can be avoided. In other words, grading in the plane could increase the operating envelope of pouch cells such that they could be fast charged without suffering from plating.

Figure \ref{fig:plating_time} shows the time when plating is initiated during a 5C charge. During this charge, the graded cell does not plate whereas the uniform electrodes experiences plating at the end of charge. The propensity of plating to occur in uniform cells at the end of charge is one of the main reasons why optimal fast charging currents are often tapered at the end of charge\cite{attia2020closed,tucker2023optimal}.

Figure \ref{fig:plating_space} illustrates the spatial distribution of plating across the plane of the uniform pouch cell at the end of the 5C charge. The grey area indicates a region where plating is predicted to occur. The figure suggests that, for these cells, most of the plating occurs opposite the tabs with the plating then progressing up the plane. This effect is in response to the current distribution being highest at the tabs at the start of the charge, causing the active particles in this region to be charged first. When these particles become fully charged, they hinder further charging in this region and so cause current to flow down towards the bottom of the pouch cell later on during the charge. By restricting  the area where the current can flow through, the local current densities in this region at the bottom of the cell are increased (see also Figure \ref{fig:comp_curr}), causing a higher likelihood of plating.




\section*{Conclusions}
A model for large format lithium-ion battery pouch cells with spatial grading across the plane was developed and validated against experimental data. An analytical solution for the optimal electrical resistance distribution across the plane of large format pouch cells to achieve a flat current distribution was derived. The main benefit of graded electrodes identified was to increase the allowable charging C-rate from  which lithium plating could be avoided; the graded electrodes increased the safe charging rate from 2.4C to 4.3C. By contrast, only a 1.2\% increase in charging capacity was achieved for 4C charging. These results indicate the potential of in-the-plane graded electrode microstructures for enabling fast charging of large format pouch cells that by limiting the spatial heterogeneities. \rd{Future work will focus on experiments to fabricate graded electrodes for LFP pouch cells and validate the results of this modelling work.}


					\footnotesize 
						
						\subsubsection*{Acknowledgements}
	Funding for this work was partially provided by the Faraday Institution project `Nextrode - next generation electrodes' (grant number FIRG015 and FIRG066). RD acknowledges funding by a UKIC Research Fellowship from the Royal Academy of Engineering.

\subsection*{Code Availability} 
The open-source code for the pouch cell simulations that have been used to produce the results of this study are available online (\url{https://github.com/EloiseTredenick/2DBatteryTemperatureGradedModel}).

\bibliographystyle{ieeetr}
\fontsize{8}{8}\rm
\bibliography{bibliog}

\begin{thebibliography}{10}

\bibitem{singh2008intercalation}
G.~K. Singh, G.~Ceder, and M.~Z. Bazant, ``Intercalation dynamics in
  rechargeable battery materials: {G}eneral theory and phase-transformation
  waves in {LiFePO$_4$},'' {\em Electrochimica Acta}, vol.~53, no.~26,
  pp.~7599--7613, 2008.

\bibitem{tredenick2024multilayer}
E.~C. Tredenick, S.~W.~R. Drummond, Y.~S. Y, S.~R. Duncan, and P.~S. Grant, ``A
  multilayer {Doyle-Fuller-Newman} model to optimise the rate performance of
  bilayer cathodes in {L}i ion batteries,'' 2024.
\newblock pre-print.

\bibitem{ho20213d}
A.~S. Ho, D.~Y. Parkinson, D.~P. Finegan, S.~E. Trask, A.~N. Jansen, W.~Tong,
  and N.~P. Balsara, ``3d detection of lithiation and lithium plating in
  graphite anodes during fast charging,'' {\em ACS Nano}, vol.~15, no.~6,
  pp.~10480--10487, 2021.

\bibitem{lauro2023restructuring}
S.~N. Lauro, J.~N. Burrow, and C.~B. Mullins, ``Restructuring the lithium-ion
  battery: {A} perspective on electrode architectures,'' {\em eScience},
  p.~100152, 2023.

\bibitem{sukenik2023impact}
E.~G. Sukenik, L.~Kasaei, and G.~G. Amatucci, ``Impact of gradient porosity in
  ultrathick electrodes for lithium batteries,'' {\em Journal of Power
  Sources}, vol.~579, p.~233327, 2023.

\bibitem{chowdhury2021simulation}
R.~Chowdhury, A.~Banerjee, Y.~Zhao, X.~Liu, and N.~Brandon, ``Simulation of
  bi-layer cathode materials with experimentally validated parameters to
  improve ion diffusion and discharge capacity,'' {\em Sustainable Energy \&
  Fuels}, vol.~5, no.~4, pp.~1103--1119, 2021.

\bibitem{cheng2022extending}
C.~Cheng, R.~Drummond, S.~R. Duncan, and P.~S. Grant, ``Extending the
  energy-power balance of {L}i-ion batteries using graded electrodes with
  precise spatial control of local composition,'' {\em Journal of Power
  Sources}, vol.~542, p.~231758, 2022.

\bibitem{cheng2020combining}
C.~Cheng, R.~Drummond, S.~R. Duncan, and P.~S. Grant, ``Combining composition
  graded positive and negative electrodes for higher performance {L}i-ion
  batteries,'' {\em Journal of Power Sources}, vol.~448, p.~227376, 2020.

\bibitem{chen2024tortuosity}
Z.~Chen and Y.~Zhao, ``Tortuosity estimation and microstructure optimization of
  non-uniform porous heterogeneous electrodes,'' {\em Journal of Power
  Sources}, vol.~596, p.~234095, 2024.

\bibitem{lee2021multi}
S.~H. Lee, C.~Huang, and P.~S. Grant, ``Multi-layered composite electrodes of
  high power {Li$_4$Ti$_5$O$_{12}$} and high capacity sno$_2$ for smart lithium
  ion storage,'' {\em Energy Storage Materials}, vol.~38, pp.~70--79, 2021.

\bibitem{yari2020constructive}
S.~Yari, H.~Hamed, J.~D’Haen, M.~K. Van~Bael, F.~U. Renner, A.~Hardy, and
  M.~Safari, ``Constructive versus destructive heterogeneity in porous
  electrodes of lithium-ion batteries,'' {\em ACS Applied Energy Materials},
  vol.~3, no.~12, pp.~11820--11829, 2020.

\bibitem{boyce2021design}
A.~M. Boyce, D.~J. Cumming, C.~Huang, S.~P. Zankowski, P.~S. Grant, D.~J.
  Brett, and P.~R. Shearing, ``Design of scalable, next-generation thick
  electrodes: opportunities and challenges,'' {\em ACS Nano}, vol.~15, no.~12,
  pp.~18624--18632, 2021.

\bibitem{tredenick2023bridging}
E.~C. Tredenick, A.~M. Boyce, S.~Wheeler, J.~Li, Y.~Sun, R.~Drummond, S.~R.
  Duncan, P.~S. Grant, and P.~R. Shearing, ``Bridging the gap between
  microstructurally resolved computed tomography-based and homogenised
  {Doyle-Fuller-Newman} models for lithium-ion batteries,'' 2023.
\newblock pre-print.

\bibitem{wu2022gradient}
J.~Wu, Z.~Ju, X.~Zhang, A.~C. Marschilok, K.~J. Takeuchi, H.~Wang, E.~S.
  Takeuchi, and G.~Yu, ``Gradient design for high-energy and high-power
  batteries,'' {\em Advanced Materials}, vol.~34, no.~29, p.~2202780, 2022.

\bibitem{drummond2022modelling}
R.~Drummond, C.~Cheng, P.~S. Grant, and S.~R. Duncan, ``Modelling the impedance
  response of graded {LiFePO$_4$} cathodes for {L}i-ion batteries,'' {\em
  Journal of the Electrochemical Society}, vol.~169, no.~1, p.~010528, 2022.

\bibitem{garrick2023atoms}
T.~R. Garrick, Y.~Zeng, J.~B. Siegel, and V.~R. Subramanian, ``From atoms to
  wheels: {T}he role of multi-scale modeling in the future of transportation
  electrification,'' {\em Journal of the Electrochemical Society}, vol.~170,
  no.~11, p.~113502, 2023.

\bibitem{ramadesigan2010optimal}
V.~Ramadesigan, R.~N. Methekar, F.~Latinwo, R.~D. Braatz, and V.~R.
  Subramanian, ``Optimal porosity distribution for minimized ohmic drop across
  a porous electrode,'' {\em Journal of the Electrochemical Society}, vol.~157,
  no.~12, p.~A1328, 2010.

\bibitem{li2023enable}
S.~Li, M.~W. Marzook, C.~Zhang, G.~J. Offer, and M.~Marinescu, ``How to enable
  large format 4680 cylindrical lithium-ion batteries,'' {\em Applied Energy},
  vol.~349, p.~121548, 2023.

\bibitem{chu2020parameterization}
H.~N. Chu, S.~U. Kim, S.~K. Rahimian, J.~B. Siegel, and C.~W. Monroe,
  ``Parameterization of prismatic lithium--iron--phosphate cells through a
  streamlined thermal/electrochemical model,'' {\em Journal of Power Sources},
  vol.~453, p.~227787, 2020.

\bibitem{lin2022multiscale}
J.~Lin, H.~N. Chu, D.~A. Howey, and C.~W. Monroe, ``Multiscale coupling of
  surface temperature with solid diffusion in large lithium-ion pouch cells,''
  {\em Communications Engineering}, vol.~1, no.~1, p.~1, 2022.

\bibitem{mikheenkova2024visualizing}
A.~Mikheenkova, A.~Sch{\"o}kel, A.~J. Smith, I.~Ahmed, W.~R. Brant, M.~J.
  Lacey, and M.~Hahlin, ``Visualizing ageing-induced heterogeneity within large
  prismatic lithium-ion batteries for electric cars using diffraction
  radiography,'' {\em Journal of Power Sources}, vol.~599, p.~234190, 2024.

\bibitem{fordham2023correlative}
A.~Fordham, Z.~Milojevic, E.~Giles, W.~Du, R.~E. Owen, S.~Michalik, P.~A.
  Chater, P.~K. Das, P.~S. Attidekou, S.~M. Lambert, {\em et~al.},
  ``Correlative non-destructive techniques to investigate aging and orientation
  effects in automotive {L}i-ion pouch cells,'' {\em Joule}, vol.~7, no.~11,
  pp.~2622--2652, 2023.

\bibitem{yari2022non}
S.~Yari, M.~K. Van~Bael, A.~Hardy, and M.~Safari, ``Non-uniform distribution of
  current in plane of large-area lithium electrodes,'' {\em Batteries \&
  Supercaps}, vol.~5, no.~10, p.~e202200217, 2022.

\bibitem{yang2019asymmetric}
X.-G. Yang, T.~Liu, Y.~Gao, S.~Ge, Y.~Leng, D.~Wang, and C.-Y. Wang,
  ``Asymmetric temperature modulation for extreme fast charging of lithium-ion
  batteries,'' {\em Joule}, vol.~3, no.~12, pp.~3002--3019, 2019.

\bibitem{liu2010visualization}
J.~Liu, M.~Kunz, K.~Chen, N.~Tamura, and T.~J. Richardson, ``Visualization of
  charge distribution in a lithium battery electrode,'' {\em The Journal of
  Physical Chemistry Letters}, vol.~1, no.~14, pp.~2120--2123, 2010.

\bibitem{bason2022non}
M.~G. Bason, T.~Coussens, M.~Withers, C.~Abel, G.~Kendall, and P.~Kr{\"u}ger,
  ``Non-invasive current density imaging of lithium-ion batteries,'' {\em
  Journal of Power Sources}, vol.~533, p.~231312, 2022.

\bibitem{mohammadi2019diagnosing}
M.~Mohammadi, E.~V. Silletta, A.~J. Ilott, and A.~Jerschow, ``Diagnosing
  current distributions in batteries with magnetic resonance imaging,'' {\em
  Journal of Magnetic Resonance}, vol.~309, p.~106601, 2019.

\bibitem{wang2024plane}
Z.~Wang, D.~L. Danilov, R.-A. Eichel, and P.~H. Notten, ``About the in-plane
  distribution of the reaction rate in lithium-ion batteries,'' {\em
  Electrochimica Acta}, vol.~475, p.~143582, 2024.

\bibitem{yazdanpour2014distributed}
M.~Yazdanpour, P.~Taheri, A.~Mansouri, and M.~Bahrami, ``A distributed
  analytical electro-thermal model for pouch-type lithium-ion batteries,'' {\em
  Journal of the Electrochemical Society}, vol.~161, no.~14, p.~A1953, 2014.

\bibitem{campillo2017effect}
J.~M. Campillo-Robles, X.~Artetxe, and K.~del Teso~S{\'a}nchez, ``Effect of
  thickness on the maximum potential drop of current collectors,'' {\em Applied
  Physics Letters}, vol.~111, no.~9, 2017.

\bibitem{cho2023improving}
C.~Cho, S.~Kelley, J.~G. Tylka, M.~He, N.~N. Nandola, and C.~D. Rahn,
  ``Improving nonuniform utilization of {L}i-ion pouch cells using tapered
  electrodes through calendering,'' in {\em International Design Engineering
  Technical Conferences and Computers and Information in Engineering
  Conference}, vol.~87301, American Society of Mechanical Engineers, 2023.

\bibitem{taheri2014theoretical}
P.~Taheri, A.~Mansouri, M.~Yazdanpour, and M.~Bahrami, ``Theoretical analysis
  of potential and current distributions in planar electrodes of lithium-ion
  batteries,'' {\em Electrochimica Acta}, vol.~133, pp.~197--208, 2014.

\bibitem{newman1993potential}
J.~Newman and W.~Tiedemann, ``Potential and current distribution in
  electrochemical cells: {I}nterpretation of the half-cell voltage measurements
  as a function of reference-electrode location,'' {\em Journal of the
  Electrochemical Society}, vol.~140, no.~7, p.~1961, 1993.

\bibitem{parmananda2023underpinnings}
M.~Parmananda, B.~S. Vishnugopi, H.~Garg, and P.~P. Mukherjee, ``Underpinnings
  of multiscale interactions and heterogeneities in {L}i-ion batteries:
  {E}lectrode microstructure to cell format,'' {\em Energy Technology},
  vol.~11, no.~11, p.~2200691, 2023.

\bibitem{hahn2023reduced}
Y.~Hahn, Z.~Gao, T.-T. Nguyen, and V.~Oancea, ``A reduced order model for a
  lithium-ion {3D} pouch battery for coupled thermal-electrochemical
  analysis,'' {\em Journal of Energy Storage}, vol.~70, p.~107966, 2023.

\bibitem{guo2013three}
M.~Guo, G.-H. Kim, and R.~E. White, ``A three-dimensional multi-physics model
  for a {L}i-ion battery,'' {\em Journal of Power Sources}, vol.~240,
  pp.~80--94, 2013.

\bibitem{pan2020computational}
Y.-W. Pan, Y.~Hua, S.~Zhou, R.~He, Y.~Zhang, S.~Yang, X.~Liu, Y.~Lian, X.~Yan,
  and B.~Wu, ``A computational multi-node electro-thermal model for large
  prismatic lithium-ion batteries,'' {\em Journal of Power Sources}, vol.~459,
  p.~228070, 2020.

\bibitem{aylagas2022cidemod}
R.~C. Aylagas, C.~Ganuza, R.~Parra, M.~Ya{\~n}ez, and E.~Ayerbe, ``cide{MOD}:
  {A}n open source tool for battery cell inhomogeneous performance
  understanding,'' {\em Journal of the Electrochemical Society}, vol.~169,
  no.~9, p.~090528, 2022.

\bibitem{doyle1993modeling}
M.~Doyle, T.~F. Fuller, and J.~Newman, ``Modeling of galvanostatic charge and
  discharge of the lithium/polymer/insertion cell,'' {\em Journal of the
  Electrochemical society}, vol.~140, no.~6, p.~1526, 1993.

\bibitem{santhanagopalan2006review}
S.~Santhanagopalan, Q.~Guo, P.~Ramadass, and R.~E. White, ``Review of models
  for predicting the cycling performance of lithium ion batteries,'' {\em
  Journal of Power Sources}, vol.~156, no.~2, pp.~620--628, 2006.

\bibitem{hosseinzadeh2018impact}
E.~Hosseinzadeh, J.~Marco, and P.~Jennings, ``The impact of multi-layered
  porosity distribution on the performance of a lithium ion battery,'' {\em
  Applied Mathematical Modelling}, vol.~61, pp.~107--123, 2018.

\bibitem{frith2023non}
J.~T. Frith, M.~J. Lacey, and U.~Ulissi, ``A non-academic perspective on the
  future of lithium-based batteries,'' {\em Nature communications}, vol.~14,
  no.~1, p.~420, 2023.

\bibitem{trefethen2000spectral}
L.~N. Trefethen, {\em Spectral methods in {MATLAB}}.
\newblock SIAM, 2000.

\bibitem{Matlab2022}
MATLAB, {\em version 9.13 (R2022b)}.
\newblock Natick, Massachusetts: The MathWorks Inc., 2022.

\bibitem{o2022lithium}
S.~E. O'Kane, W.~Ai, G.~Madabattula, D.~Alonso-Alvarez, R.~Timms, V.~Sulzer,
  J.~S. Edge, B.~Wu, G.~J. Offer, and M.~Marinescu, ``Lithium-ion battery
  degradation: how to model it,'' {\em Physical Chemistry Chemical Physics},
  vol.~24, no.~13, pp.~7909--7922, 2022.

\bibitem{reniers2019review}
J.~M. Reniers, G.~Mulder, and D.~A. Howey, ``Review and performance comparison
  of mechanical-chemical degradation models for lithium-ion batteries,'' {\em
  Journal of the Electrochemical Society}, vol.~166, no.~14, pp.~A3189--A3200,
  2019.

\bibitem{couto2021faster}
L.~D. Couto, R.~Romagnoli, S.~Park, D.~Zhang, S.~J. Moura, M.~Kinnaert, and
  E.~Garone, ``Faster and healthier charging of lithium-ion batteries via
  constrained feedback control,'' {\em IEEE Transactions on Control Systems
  Technology}, vol.~30, no.~5, pp.~1990--2001, 2021.

\bibitem{romagnoli2019feedback}
R.~Romagnoli, L.~D. Couto, A.~Goldar, M.~Kinnaert, and E.~Garone, ``A feedback
  charge strategy for {L}i-ion battery cells based on reference governor,''
  {\em Journal of Process Control}, vol.~83, pp.~164--176, 2019.

\bibitem{attia2020closed}
P.~M. Attia, A.~Grover, N.~Jin, K.~A. Severson, T.~M. Markov, Y.-H. Liao, M.~H.
  Chen, B.~Cheong, N.~Perkins, Z.~Yang, {\em et~al.}, ``Closed-loop
  optimization of fast-charging protocols for batteries with machine
  learning,'' {\em Nature}, vol.~578, no.~7795, pp.~397--402, 2020.

\bibitem{tucker2023optimal}
G.~Tucker, R.~Drummond, and S.~R. Duncan, ``Optimal fast charging of lithium
  ion batteries: Between model-based and data-driven methods,'' {\em Journal of
  the Electrochemical Society}, vol.~170, no.~12, p.~120508, 2023.

\bibitem{Kashkooli2016}
A.~G. Kashkooli, S.~Farhad, D.~U. Lee, K.~Feng, S.~Litster, S.~K. Babu, L.~Zhu,
  and Z.~Chen, ``Multiscale modeling of lithium-ion battery electrodes based on
  nano-scale x-ray computed tomography,'' {\em Journal of Power Sources},
  vol.~307, pp.~496--509, 3 2016.

\end{thebibliography}
 
\section*{Appendix}

\renewcommand{\thefigure}{A\arabic{figure}}
\renewcommand{\theequation}{A\arabic{equation}}
\renewcommand{\thetable}{A\arabic{table}}
\setcounter{figure}{0} 
\setcounter{equation}{0} 
\setcounter{table}{0} 




The open circuit voltage (OCV) function for the LFP cells\cite{Kashkooli2016,lin2022multiscale} at 0.02 C is modelled by:   
\begin{multline}
	U (\text{SoC},T) = \frac{dU}{dT}  \ \left(T-T_{\scriptscriptstyle \text{ref}} \right)    +0.0027 \ T \ I_\text{sign}+  3.382+0.0047 \ \text{SoC}+1.627 \ \exp(-81.163 \ \text{SoC}^{1.0138})\\+7.6445\times 10^{-8} \ \exp(25.36 \ \text{SoC}^{2.469}) 	- 8.441\times 10^{-8} \ \exp(25.262 \ \text{SoC}^{2.478}) - 3.3852+3.2585 ,\label{eq66} 
\end{multline}
where 
$dU/dT =  -10^{-4}$.
	 The initial conditions for the simulations are: 
	 \begin{subequations}
	 	\begin{align}
   	\text{SoC}(0) &= \text{SoC}_0 ,  \  \text{on} \ \Omega\  \text{and} \  \partial\Omega, \label{IC3} \\
	 		U(0) &= U_0 = U(\text{SoC}_0),  \  \text{on} \ \Omega\  \text{and} \  \partial\Omega, \label{IC5} \\
	 		V(y,z,0) &= V_0 = U_0 +   \ R_0  \ I_{\scriptscriptstyle \text{app}}  ,  \  \text{on} \ \Omega\  \text{and} \  \partial\Omega, \label{IC1} \\
     		v_1(y,z,0) &= 0 ,  \  \text{on} \ \Omega\  \text{and} \  \partial\Omega, \label{IC3} \\
          		v_2(y,z,0) &= 0 ,  \  \text{on} \ \Omega\  \text{and} \  \partial\Omega, \label{IC4} \\
	 		T(y,z,0)  &= T_{\scriptscriptstyle \text{ref}} = T_0 ,  \  \text{on} \ \Omega\  \text{and} \  \partial\Omega. \label{IC2} 
	 	\end{align}
	 \end{subequations}

	 \begin{table} [h!]
	 	\footnotesize
	 	\centering
	 	\caption[Variables]{Variables of the model. }	 	\label{VariablesModel}	
	 	\begin{tabular}{    p{1.2cm}  p{5cm}  p{1cm} } 
	 		\hline
	 		Variable & Description  & Unit   \\ \hline 
	 			$i_x$	 & Current  density through electrode thickness & A m$^{-2}$  \\
     	 			$I$	 & Current  density in the current collector & A m$^{-2}$  \\
     	 			$I_{\scriptscriptstyle \text{app}}$	 & Applied current & A  \\ 
                    $I_\text{sign}$	 & Sign of the current &   \\
     	$R$ &   Cell series resistance & $\Omega$   \\ 
	 		SoC	 &   Local state of charge  &  -  \\ 
	 			$T$	 & Temperature &  K  \\ 
     	$t$ 	 &   Time  &  s    \\ 
	 		$U$ 	 &  Open circuit voltage (OCV)     &   V    \\ 
	 		$V$ 	 & Voltage    &   V   \\ 
    	 		$v_{1}$ 	 & Voltage  across RC-pair  &   V   \\ 
        	 		$v_{2}$ 	 & Voltage  across RC-pair   &   V   \\ 
	 		$x$	 &  Direction through cell   &  m    \\ 
	 		$y$	 &  Direction along cell width   &  m    \\ 
	 		$z$	 &  Direction along cell height   &  m    \\
	 		\hline 
	 	\end{tabular}
      \label{tab:vars}
	 \end{table}

	 \begin{table} [h!]
	 	\scriptsize
	 	\centering
	 	\caption[Model Parameters]{Model parameters. Experimental values obtained from \cite{lin2022multiscale} for an 80Ah capacity cell.}	 	\label{ParametersModel}	
	 	\begin{tabular}{    p{1.3cm}  p{7.0cm}  p{1.5cm}  p{2.2cm}  p{1.3cm} } 
	 		\hline
	 		Parameter & Description  & Units  & Value & Source   \\ \hline 
	 		$A_{\scriptscriptstyle \text{tab}}$& Area of a tab& m$^2$ & 1.56$\times 10^{-4}$ & Experiment   \\
    	 		$A_{\scriptscriptstyle \text{cell}}$ & Cell area $(L_y \times L_z)$ & m$^2$ & $30 \times 10^{-3}$ &  Experiment  \\
            	 		$n_\text{layers}$ & Number of electrode layers in cell &  &42 &  Experiment  \\
	 			$C_{\scriptscriptstyle \text{cell}}$& Capacity of cell (20~Ah) per cell area   &  A s   & 20 $\times 3600 $ &  Experiment   \\ 		 
     $C_1$&  Capacitance of RC-pair & F& 2.79$\times 10^{4}$&  Fitting\\ 
 		$C_2$&  Capacitance of RC-pair & F  &   8.89$\times 10^{3}$  &  Fitting  \\ 
    	 	
	 		$h$ & Heat transfer coefficient & W m$^{-2}$ K$^{-1}$ & 13.2 &      Fitting\\ 	
     		$ h_{\scriptscriptstyle \text{tab}} $ & Heat transfer coefficient of tabs & W m$^{-2}$ K$^{-1}$ &  
   51.58 &     Fitting \\	
            		$  h_{\scriptscriptstyle \text{faces}} $ & \rd{Effective} heat transfer coefficient of current collector faces & W m$^{-2}$ K$^{-1}$ & 210$ \times L_x$  &    Fitting  \\	
  $L_{\scriptscriptstyle \text{cn}}$& Thickness of negative current collector in $x$ direction & $\mu$m &  $25$   &  Experiment   \\ 
  $L_{\scriptscriptstyle \text{cp}}$& Thickness of positive current collector in $x$ direction & $\mu$m &  $25$   &  Experiment   \\    
   $L_{\scriptscriptstyle \text{x}}$& Electrode thickness  &  m  & 110 $\times 10^{-6}$  & Experiment   \\ 
	 		$L_{\scriptscriptstyle \text{y}}$& Height of battery & m & $0.2$ & Experiment   \\ 		
	 		$L_{\scriptscriptstyle \text{z}}$& Width of battery & m & $0.15$ & Experiment   \\ 	
 	$R_0$&  Series resistance of Uniform cell &$\Omega$  &1.5$\times 10^{-3} $&  Fitting \\ 
	 		$R_1$&  Resistance of RC-pair & $\Omega$  &  1.10$\times 10^{-3}$&  Fitting\\ 
        	 		$R_2$&  Resistance of RC-pair &$\Omega$  &2.25$\times 10^{-4} $&   Fitting \\ 
                    	 		$T_{\scriptscriptstyle \text{ref}}$	&	Constant absolute reference temperature& K & 298.15 ($25\,^{\circ}\mathrm{C}$)& Experiment   \\ 
	 			$\lambda_{\scriptscriptstyle \text{eff}}$& Effective thermal conductivity  & W m$^{-1}$ K$^{-1}$ & 4.5 &   Fitting   \\ 
	 			 		$\rho$& Volumetric heat capacity & J K$^{-1}$ m$^{-3}$ & 1.35 $\times 10^5$ &   Fitting   \\ 
   $\sigma_{\scriptscriptstyle \text{cn}}$& Electronic conductivity of copper \rd{negative} current collector & S m$^{-1}$ &   $5.96\times10^{7}$  &  Experiment   \\ 
	 		$\sigma_{\scriptscriptstyle \text{cp}}$&  Electronic conductivity of aluminium \rd{positive} current collector & S m$^{-1}$ &$ 3.77\times10^{7}$& \\
	 		\hline 
	 	\end{tabular}
   \label{tab:params}
	 \end{table}

\begin{figure} [h!]
	\hspace{-50pt}
\includegraphics[width=0.82\textheight,keepaspectratio]{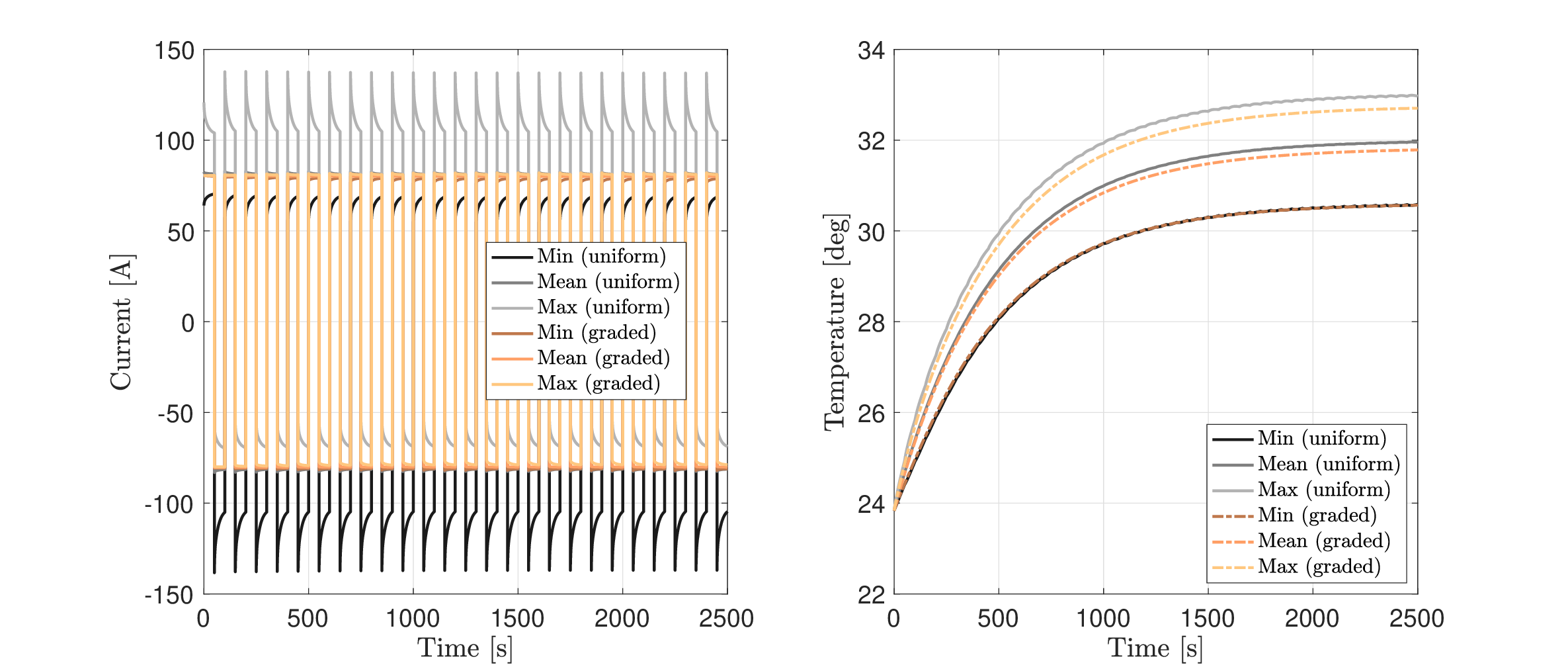}
\caption{Comparison between the graded and uniform models during cycling at 4C and an applied current of 80~A. The current and temperature responses with time are shown, including the minimum, average, and maximum temperatures across the plane of the pouch cell.
}
\label{fig:GradedvsUniformcycled}  
\end{figure}

\end{document}